\documentclass[aps,twocolumn,prb,showpacs,floatfix,amsmath,amssymb,superscriptaddress]{revtex4-1}
\usepackage{graphicx}
\usepackage{bm}
\usepackage{hyperref}
\usepackage{color}
\usepackage{natbib}
\usepackage{changes,cancel,amsmath, amssymb}
\usepackage{braket,float,soul}
\usepackage{comment}
\usepackage{ulem}
\usepackage{graphicx}
\usepackage{bm}
\usepackage{amsmath,amssymb}
\usepackage{braket,float,soul}
\usepackage{comment}
\definecolor{ao(english)}{rgb}{0.0, 0.5, 0.0}
\begin{document}
\excludecomment{taglio}
\title{Long time rigidity to flux-induced  symmetry breaking  in quantum quench dynamics}

\author{Lorenzo  Rossi}

\affiliation{Dipartimento di Scienza Applicata e Tecnologia, Politecnico di Torino, 10129 Torino, Italy}

\author{Luca Barbiero}
\affiliation{Dipartimento di Scienza Applicata e Tecnologia, Politecnico di Torino, 10129 Torino, Italy}

\author{Jan Carl Budich}
\affiliation{Institute of Theoretical Physics, Technische Universit\"at Dresden and
W\"urzburg-Dresden Cluster of Excellence ct.qmat, 01062 Dresden, Germany}

\author{Fabrizio Dolcini}
\email{fabrizio.dolcini@polito.it}
\affiliation{Dipartimento di Scienza Applicata e Tecnologia, Politecnico di Torino, 10129 Torino, Italy}

\begin{abstract}
We investigate how the breaking of charge conjugation symmetry $\mathcal{C}$ impacts on the dynamics of a half-filled fermionic lattice system  after global quenches. We show that, when the initial state is insulating and the $\mathcal{C}$-symmetry is broken non-locally by a constant magnetic flux, local observables and correlations behave as if the symmetry were unbroken for a time interval proportional to  the system size $L$.   In particular,   the local particle density of a quenched dimerized insulator remains pinned to $1/2$ in each lattice site for an extensively long  time, while it starts to significantly fluctuate only afterwards.  Due to its qualitative resemblance to the sudden arrival of rapidly rising ocean waves, we dub this  phenomenon   the ``tsunami effect".  Notably,  it  occurs even though the chiral symmetry  is  dynamically broken right after the quench. Furthermore, we identify a way to quantify the amount of symmetry breaking in the quantum state, showing that  in insulators perturbed by a flux  it is exponentially suppressed as a function of the system size, while it is only algebraically suppressed in metals and in insulators with locally broken $\mathcal{C}$-symmetry. The robustness of the tsunami effect to weak disorder and interactions is demonstrated, and possible experimental realizations are proposed.
\end{abstract}

\maketitle
\section{Introduction}   
One of the most surprising aspects of quantum mechanics is the non-local  effect induced by a magnetic flux. Indeed, in striking contrast to classical particles, a charged quantum particle  can experience the presence of a flux even in regions where no magnetic field is   present. The  Aharonov-Bohm effect\cite{aharonov-bohm,peshkin-lipkin} is a paradigmatic manifestation of this phenomenon.
Nevertheless, metals and   insulators  are known to exhibit a quite different response to a magnetic flux, encoded within the linear response theory in the Drude weight~$D$.\cite{kohn_1964,shastry_PRL_1990,kawakami_PRB_1991,zotos_1995} Such a quantity, which is obtained from  equilibrium state correlation functions,  describes the current generated by a flux-induced electric pulse. While $D$ is finite for metals, it vanishes  for an insulator, implying that the ground state energy of an insulator is insensitive to  magnetic flux variations.
It has also been proven that, as long as flux changes adiabatically, such insensitivity to a flux holds also beyond linear response theory, i.e. in higher order Drude weigths.\cite{oshikawa1,oshikawa2}  In this respect, insulators exhibit a more ``classical behavior" than metals, in view of the localized character of their correlations.\cite{kohn_1964}

In recent years, the experimental advances in cold atom setups\cite{dalibard-review_2011,spielman_RPP_2014,goldman-budich-zoller,spielman_RMP_2019}  have motivated an intense theoretical activity to investigate the effects of a flux  in   far away from equilibrium settings, like in quantum quenches\cite{Calabrese_PRL_2006,Polkovnikov_RMP_2011,Eisert_NP_2015,Mitra_ARCMP_2018}, where the adiabaticity condition does not hold. Most works have analyzed  flux quenches in fermionic  ring-shaped lattice systems, where a flux   is suddenly switched on or off, mainly with the purpose of determining whether a persistent current can flow in these conditions,  how it is possibly affected by interactions, and what is its nonlinear dependence on the flux.\cite{deluca_2014,misguich,rossi-dolcini_2022}

It should be pointed out, however, that  any  flux quench protocol  induces a {\it local}  electric pulse, which is thus a classical effect {\it per se}. In order to probe genuine  quantum mechanical  non-local effects of the flux in the far from equilibrium dynamics, it is important to rather consider  a constant magnetic flux, and to quench   other parameters. In these conditions, a crucial question is still open: When an insulator is driven out of equilibrium, how long does it remain insensitive to the presence of a flux? 
\begin{figure}[h]
\centering
\includegraphics[width=  \columnwidth]{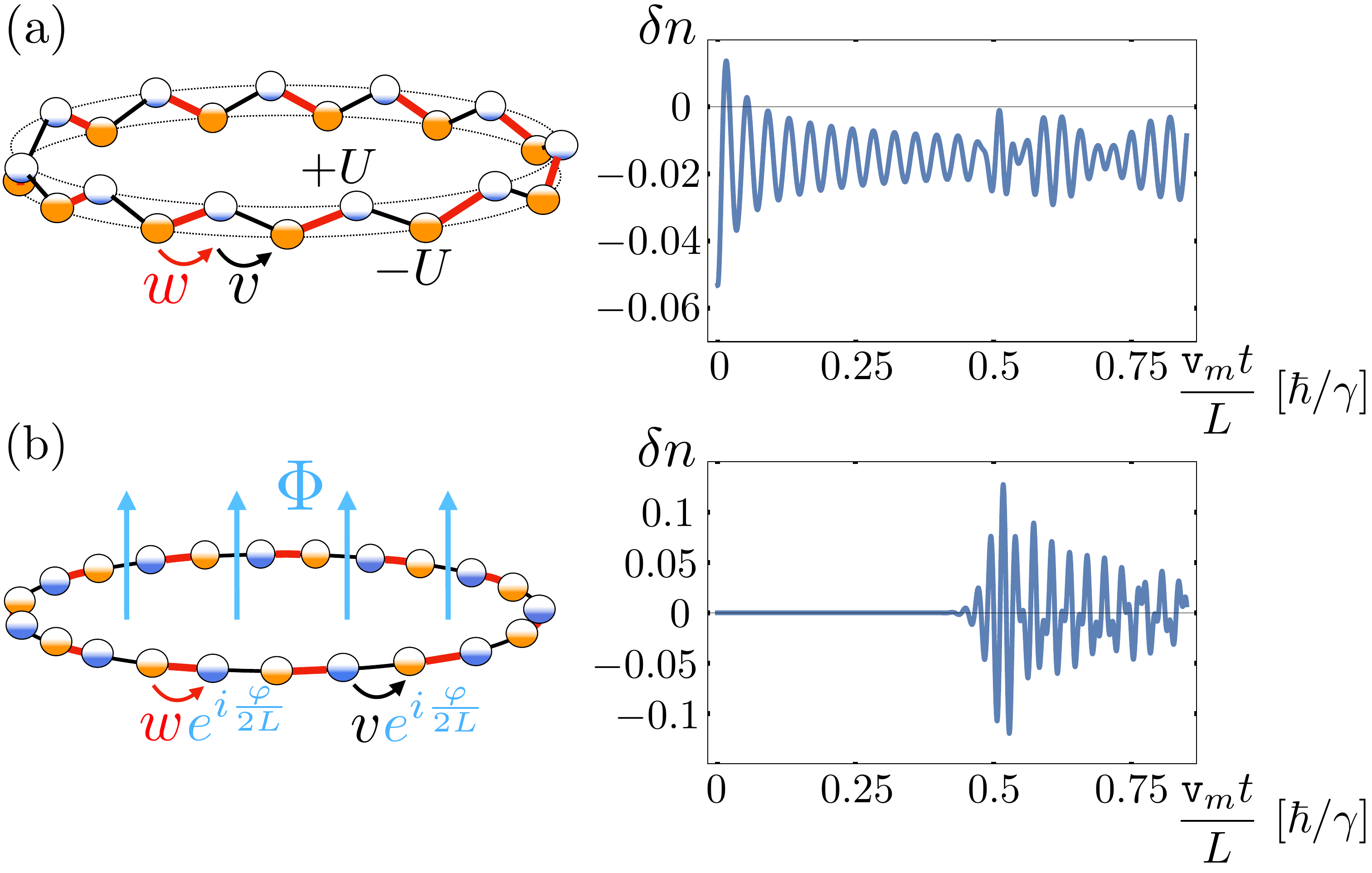}
\caption{\label{Fig1} Quantum dynamics  resulting from a quench applied to a dimerized ring-shaped model (\ref{Ham-real}) with $L=80$ cells (160 sites), where   the tunneling amplitudes, with initial values  $v^i =1, w^i=0.5 $,   are exchanged $w \leftrightarrow v$. The deviation $\delta n$ from 1/2 of the on-site density  is shown as a function of time.  
(a) the case of a staggered potential  $U=1/10$    breaking the charge conjugation symmetry $\mathcal{C}$  locally. (b) the case of a non-local $\mathcal{C}$-symmetry breaking induced by a flux  phase  $\varphi=\pi/2$,  where $\varphi=2\pi\Phi/\Phi_0$, with $\Phi_0$ denoting the flux quantum.}
\end{figure}

Interestingly, this type of problems can also be considered from a different perspective, namely the impact of symmetry breaking.  
In quantum systems, where the state is characterized by a density matrix operator $\boldsymbol\rho$ and its evolution is governed by the Hamiltonian ${H}$, one has to consider   the effect of symmetries on both  $\boldsymbol\rho$ and ${H}$, as they are not necessarily related.
Indeed a Hamiltonian may fulfill a given symmetry, despite each of its degenerate ground states may not, as is the case for the spontaneous symmetry breaking.\cite{landau_book,wen_book} Conversely, the ground state  of a Hamiltonian may fulfill a symmetry, despite ${H}$ itself does not, because of other excited states.

Although the existence of a symmetry is per se a ``binary"  concept --it is either  fulfilled or not-- one may wonder whether and to what extent   the effects of a symmetry breaking become observable.  A flux, for instance, breaks time reversal and charge-conjugation symmetries in the Hamiltonian.\cite{review_NJP,review_RMP}  Yet, the rigidity of  a band insulator to the flux can also be regarded as an example where the  ground state  ``does not see" these symmetry breaking, and it behaves as if the symmetries were fulfilled.
 When it comes to out of equilibrium conditions, the effect of symmetries becomes more subtle. It is known, for instance, that a symmetry can get dynamically broken even when it is fulfilled by both the initial state and the evolution Hamiltonian.\cite{Cooper_PRL_2018,Cooper_PRB_2019}. In   general terms, one can thus formulate  the following problem: When a fermionic   system with a broken symmetry is driven far from equilibrium, how and when will the effects  of such symmetry breaking become observable?

 This paper is devoted to address the above questions. Specifically,  focussing on the  breaking of the charge-conjugation symmetry $\mathcal{C}$, we analyze the quantum dynamics of a  half-filled fermionic  model in a 1D ring-shaped dimerized lattice, where a quench in the magnitude of the tunneling amplitudes is performed.
We show that,  when $\mathcal{C}$-symmetry is  broken, the physical behavior at local level crucially depends on the i)  the localization properties of the quantum state and ii) the nature of the symmetry breaking (local {\it vs} non-local). In particular, if the initial state  exhibits power-law decay correlations, as is the case of a metal, the effects of $\mathcal{C}$-symmetry breaking emerge at any time, regardless of whether the symmetry is broken locally or non-locally. 

However, if the quantum state is characterized by exponentially decaying correlations, as is the case of an insulator, a twofold scenario can emerge. A local breaking of the   $\mathcal{C}$-symmetry    straightforwardly impacts on local observables right after the quench. In contrast, if the  $\mathcal{C}$-symmetry is broken non-locally,   the  expectation values of local observables   remain  effectively pinned to the unbroken symmetry case for a   time that scales linearly with the system size. 
This result is highlighted in Fig.\ref{Fig1}, which shows the   dynamical behavior of a half-filled quenched Su-Schrieffer-Heeger (SSH) model\cite{SSH_PRL1979,SSH_PRB1980}, comparing the cases where $\mathcal{C}$-symmetry is broken  locally by an on-site staggered potential [panel (a)], and  non-locally by a magnetic flux [panel (b)].  While in the former case the site density deviates by $\delta n$ from the uniform half-filled value $1/2$ right after the quench, in the latter scenario the deviation  remains exponentially small in the system size  $L$ for a time that is proportional to $L$. Only after such a time, the density starts to wildly deviate from the value $1/2$ and to oscillate. This phenomenon, which we dub tsunami effect, manifests also in the particle current and in the evolution of correlation functions. Notably, such an extensively long time rigidity of these quantities occurs in spite of the  chiral symmetry of the SSH model, which gets dynamically broken immediately after the quench, as we   quantitatively prove. 

These results represent one further step in the exploration of the space-time scaling regime, which investigates  the out of equilibrium dynamics  for timescales that scale linearly with the  system size.
While in Ref.[\onlinecite{rossi-budich-dolcini}]  this regime was recently shown to provide a new way to topologically classify out of equilibrium  quantum systems, the present work highlights that, when an insulator  is driven out of equilibrium,  the quantum non locality induced by a flux  becomes observable in such a regime. Indeed, as we shall discuss, the tsunami effect can also be regarded as a   {\it dynamical crossover} characterizing   the local density as a function the  ratio $\eta$ between time $t$ and system  size~$L$.

The paper is organized as follows. In Sec.\ref{sec2}, after briefly recalling the definition and properties of charge conjugation transformation $\mathcal{C}$, we identify in general terms the implications of a $\mathcal{C}$-symmetric many-particle quantum state on the single-particle density matrix. Then, in Sec.\ref{sec3} we specify the   two-band model considered here, pointing out the difference between local and non-local $\mathcal{C}$-symmetry breaking. Section \ref{sec4} is devoted to the investigation of the dynamics resulting from the quench and to the demonstration of the tsunami effect. In particular, for the SSH ring-shaped lattice threaded by a flux,  we show that  the local density, current and correlations remain rigid to the non-local $\mathcal{C}$-symmetry breaking   for an extensively long time. After providing an intuitive explanation of such effect in terms of Wannier functions, in Sec.\ref{sec5} we argue that it can be quantitatively explained by   introducing  suitable   symmetry breaking quantifiers. This enables us to quantitatively distinguish global from local  $\mathcal{C}$-symmetry breaking, and also to show  that  the phenomenon occurs  even though  the chiral symmetry is   dynamically broken immediately after the quench. Moreover, in Sec.\ref{sec6} we demonstrate the stability of the tsunami effect to weak disorder and interactions. Finally, in Sec.\ref{sec7} we draw our conclusions and outline some possible experimental realizations to observe the predicted effect.

\section{Charge conjugation symmetry}
\label{sec2}  
Let us consider a   system of spinless fermions, possibly interacting or with disorder, in a one-dimensional (1D) ring-shaped lattice with periodic boundary conditions (PBCs). The quantum state of the system, characterized by a many-particle density matrix $\boldsymbol\rho$, may be an equilibrium or an out of equilibrium state at a given time~$t$.  
Without loss of generality, we shall adopt a notation where lattice sites  are grouped ``in pairs". Each pair, labelled by an integer $j=1,2\ldots L$, contains an odd and an even site denoted as $A$ and $B$, respectively, with ${c}_{j \alpha}^\dagger,{c}^{}_{j \alpha}$ (with $\alpha=A,B$) denoting the on-site fermionic creation and annihilation operators, respectively. The reason for choosing this notation is that, although for the moment we shall be quite general, later below we shall focus on dimerized lattices, and each pair   will thus be identified as a cell that is periodically repeated in the lattice.

\subsection{Charge conjugation}
\label{sec2a}
Charge conjugation $\mathcal{C}_\theta$ is a linear unitary transformation  ($\mathcal{C}^{}_\theta i\mathcal{C}_\theta^\dagger = i$ and $\mathcal{C}_\theta^\dagger=\mathcal{C}_\theta^{-1}$)  locally mapping particles into holes as follows\cite{note-on-cc}
\begin{equation}\label{CC-def}
    \mathcal{C}^{}_\theta {c}^{\dagger}_{j \alpha} \mathcal{C}_\theta^\dagger =  (\mathsf{U}_\theta)_{j\alpha; l \beta} \, {c}^{}_{l \beta} \quad,
\end{equation}
where a summation over repeated indices is meant, and
\begin{equation}\label{U-def}
\mathsf{U}_\theta=\bigoplus_{j=1}^L \left( \begin{array}{lcl} e^{2i \theta_{j A}} & 0 \\ 0 & -e^{2i \theta_{j B}}\end{array} \right)  
\end{equation}
 is a $2L \times 2L$ block-diagonal matrix in real space, characterized by the set of phases  $\{\theta\}=\{\theta_{1 A}, \theta_{1 B}\ldots \theta_{L A}, \theta_{L B}\}$.
The minus sign appearing in Eq.(\ref{U-def}) for the even ($=B$) sites is purely conventional and can always be included in the $\theta_{jB}$-phases.\\

A many-particle quantum state   is charge conjugation symmetric  if there exists a choice of the phases  $\{\theta \}$ such that the many-particle density matrix operator $\boldsymbol\rho$ fulfills $\mathcal{C}^{}_\theta\boldsymbol\rho\mathcal{C}_\theta^\dagger=\boldsymbol\rho$. 
To identify the implications of such a symmetry,  it is useful to consider the single-particle reduced density matrix, which is the tool needed to compute any one-body expectation value (in particular observables and   equal-time correlations), and is defined as
\begin{equation}\label{SP-rho-def}
\rho_{j_1\alpha_1, j_2 \alpha_2}=\langle {c}^\dagger_{j_2 \alpha_2} {c}^{}_{j_1 \alpha_1}\rangle= {\rm Tr}[\boldsymbol\rho\, {c}^\dagger_{j_2 \alpha_2} {c}^{}_{j_1 \alpha_1}] \quad,
\end{equation}
with $\alpha_1,\alpha_2=A,B$.
The following general results, whose  proof can be found in the Appendix~\ref{AppA}, hold.\\ 

\noindent {\it Theorem.} If $\mathcal{C}^{}_\theta \boldsymbol\rho\mathcal{C}_\theta^\dagger=\boldsymbol\rho$ for a set $\{\theta \}$ of phases in. Eq.(\ref{CC-def}), then\\

i) the single-particle reduced density matrix $\rho$ in Eq.(\ref{SP-rho-def}) 
fulfills the following property
\begin{equation}\label{TH1}
 ( \rho- \mathbb{I} /2)+  \mathsf{U}_\theta^* \,  ( \rho^*- \mathbb{I} /2) \,  \mathsf{U}_\theta  = 0 \quad,
\end{equation}
where $\mathsf{U}_\theta$ is given by Eq.(\ref{U-def}) and
 $\mathbb{I}$ is the identity  in the $2L$-dimensional single-particle Hilbert space. A straightforward implication  of Eq.(\ref{TH1}), obtained by taking its trace, is that $N={\rm tr}\rho=L$, where $N$ is the number of particles, i.e. {\it half-filling is a necessary condition for the state $\boldsymbol\rho$ to be $\mathcal{C}$-symmetric}. In order to avoid trivial violations of the $\mathcal{C}$-symmetry, in the rest of our paper we shall thus consider the half-filling situation.\\

 ii)  the expectation value of the local density  is 
\begin{equation} \label{TH2}
\langle {n}_{j\alpha}\rangle \equiv 1/2 \hspace{1cm} \forall j, \quad \alpha=A,B
\end{equation}
Notice that this is a stronger implication than the overall half-filling condition, for it holds also in the presence of interactions or disorder, where translational invariance is a priori broken.\\

iii) if an initially $\mathcal{C}$-symmetric quantum state   $\boldsymbol\rho^i$ evolves according to  a $\mathcal{C}$-symmetric Hamiltonian ${H}$ ($\mathcal{C}^{}_\theta {H} \mathcal{C}^\dagger_\theta= {H}$), then the evolved state $\boldsymbol\rho(t)$, remains $\mathcal{C}$-symmetric at any time~\cite{NJP-rossi-dolcini}. This is due to the unitarity of  $\mathcal{C}$ and implies, in particular, that Eqs.(\ref{TH1}) and (\ref{TH2}) hold at any time. 
\begin{equation}\label{pinning}
\langle {n}_{j \alpha} \rangle(t) \equiv \frac{1}{2} \hspace{1cm} \forall j, t \quad \alpha=A,B\quad.
\end{equation}
In the presence of $\mathcal{C}$-symmetry, the density thus remains pinned to $1/2$ even out of equilibrium. In the following, we shall analyze whether and how Eq.(\ref{pinning}) is modified by a $\mathcal{C}$-symmetry  breaking.

\subsection{Charge conjugation and gauge transformations}
\label{sec2b}
We conclude this section by observing that the matrix~(\ref{U-def}) characterizing the charge-conjugation transformation~(\ref{CC-def}) can also be rewritten as $\mathsf{U}_\theta=\mathsf{U}_0\, \Lambda^2_\theta$, where  
\begin{equation}
\mathsf{U}_0= \oplus_j \sigma_z \label{U0-def}
\end{equation}
 is the matrix obtained  for $\{0\}=\{\theta_{j \alpha}\equiv 0 \,\, \forall j,\alpha\}$, and 
\begin{equation}\label{Lambda-theta-def}
\Lambda_\theta=\bigoplus_{j=1}^L \left( \begin{array}{lcl} e^{i  \theta_{j A}  } & 0 \\ 0 & e^{i \theta_{j B}  }  \end{array} \right)  \quad.
\end{equation}
This means that the charge-conjugation transformation $\mathcal{C}_\theta$ in Eq.(\ref{CC-def}) can be interpreted as the   $\mathcal{C}_0$ transformation 
\begin{equation}
\left\{ \begin{array}{lcl}
\mathcal{C}^{}_0\, {c}^\dagger_{jA} \,\mathcal{C}^{\dagger}_0 &=&+{c}^{}_{jA} \\ & & \\
\mathcal{C}^{}_0\, {c}^\dagger_{jB} \,\mathcal{C}^{\dagger}_0 &=&-{c}^{}_{jB} 
\end{array}\right. \quad, \label{C0-def}
\end{equation}
combined
with a  local gauge transformation 
  on the fermionic operators\cite{nota-factor-2}
\begin{equation}\label{gauge-def}
{c}^\dagger_{j\alpha}=e^{i \theta_{j\alpha} } \tilde{c}^\dagger_{j\alpha}\quad,
\end{equation}
which is compactly written as
\begin{equation}
[0] \,\, \overset{\{ \theta \}}{\rightarrow} \,\, [g] \quad, \label{0-to-g}
\end{equation}
where $[0]$ denotes the original gauge of the ${c}^\dagger$-operators and $[g]$   the new gauge of the $\tilde{c}^\dagger$-operators.
As a consequence of Eq.(\ref{gauge-def}), the entries of the single-particle density matrix~(\ref{SP-rho-def}) also change  by such phase factors, 
\begin{equation} \label{rho-gauge}
\rho \, \,\rightarrow \,\, {\rho}(\{\theta\})=\Lambda^{}_\theta \rho \Lambda^*_\theta \quad.
\end{equation} 
One can thus  reformulate the above theorem  as follows. If a quantum state $\boldsymbol\rho$ is charge-conjugation symmetric for a phase set $\{\theta\}$, then by performing the local gauge transformation 
(\ref{gauge-def})-(\ref{0-to-g})
the single-particle density matrix ${\rho}(\{ \theta\})$ in the $\tilde{c}^\dagger$-basis fulfills 
\begin{equation}\label{TH1-tilde}
 [ {\rho}(\{ \theta\})- \mathbb{I} /2]+  \mathsf{U}_0  \,  [  {\rho}^*(\{ \theta\})- \mathbb{I} /2] \,  \mathsf{U}_0    = 0 \quad.
\end{equation}
In other words, one can equivalently define the charge conjugation transformation only as $\mathcal{C}_0$ in Eq.(\ref{C0-def}), provided that the possibility to perform gauge transformations (\ref{gauge-def}) is   also included. This is   because the $\mathcal{C}_0$-symmetry  may be present in the state, but hidden by an inappropriate choice of the gauge. \\

Since expectation values of observables are independent of the choice of the gauge, in the following of the paper we shall often harness this interpretation in terms of gauge transformations.

\section{Model, local vs non-local $\mathcal{C}$-symmetry breaking} 
\label{sec3}  
\subsection{The model}
So far, the above statements have been quite general. In order to illustrate their applications we shall focus on a specific   model in a dimerized lattice  with  PBCs, described by the following Hamiltonian
\begin{eqnarray}\label{Ham-real}
    {H} &=&  \gamma\left[  \sum_{j=1}^{L}   \left( v\,e^{i \frac{\varphi}{2L} }{c}_{j A}^\dagger {c}^{}_{j B} + w \, e^{i\frac{\varphi}{2L} }{c}_{j B}^\dagger {c}^{}_{j+1 A} + \text{H.c.} \right) +  \right. \nonumber \\ 
    & &  \hspace{0.3cm} \left. +  U\sum_{j=1}^L   \left(  {n}_{j A} -  {n}_{j B}  \right)  \right]\quad.
    \end{eqnarray}
The first line in Eq.(\ref{Ham-real})  is the SSH model\cite{SSH_PRL1979,SSH_PRB1980}, where $\gamma$ is an  energy scale unit related to the bandwidth,   $v$ and $w$ are the (real) dimensionless intra-cell and inter-cell hopping amplitudes, respectively, while  $\varphi=2\pi \Phi/\Phi_0$ is the phase related to the magnetic flux~$\Phi$ threading the ring, in units of the flux quantum $\Phi_0$. The second line describes an on-site staggered potential $U$ in the two cell sites, in units of~$\gamma$. For $U\neq 0$, the full model (\ref{Ham-real}) is known as the  Rice-Mele (RM) model.\cite{rice-mele}  \\

Let us recall a few aspects of the model Eq.(\ref{Ham-real}) that will be necessary to illustrate   the gist of the tsunami effect. Exploiting the cell translational invariance, the Hamiltonian Eq.(\ref{Ham-real}) can be suitably  rewritten in  momentum space as
\begin{eqnarray}\label{Ham-k}
    {H}[\phi] 
    &=&  \gamma \sum_{k =-\pi}^{\pi} ({c}^{\dagger}_{k A},{c}^{\dagger}_{k B}) \,
     { \mathbf{d}(k)  \cdot \boldsymbol\sigma} 
    \begin{pmatrix} {c}^{}_{k A}\\ {c}^{}_{k B}\end{pmatrix}
\end{eqnarray}
where   the $k$ wavevectors, here measured in units of the inverse cell size, are quantized  as $k  = 2\pi \, n/L$, where $n\in \left\{ {-\left\lfloor  L/2 \right \rfloor, \ldots , \left\lfloor (L-1)/2 \right\rfloor } \right\}$. Moreover $\boldsymbol\sigma =(\sigma_x,\sigma_y, \sigma_z)$ are Pauli matrices acting on the sublattice degree of freedom, and 
\begin{eqnarray}
 \mathbf{d}(k)  \, \,   
&= &   (  v \cos\frac{\varphi}{2L} + w \cos(k +\frac{\varphi}{2L})   ,  \nonumber  \\
& & -v \sin\frac{\varphi}{2L} + w \sin(k +\frac{\varphi}{2L}) ,     U  )  \label{d-vec}
\end{eqnarray}
denotes the vector field along the Brillouin zone. The spectrum consists of two bands $E_\pm(k)=\gamma \varepsilon_\pm(k)$, where the dimensionless dispersion relations $\varepsilon_\pm(k)=\pm |\mathbf{d}(k)|$   explicitly read 
\begin{equation}\label{spectrum}
\varepsilon_\pm(k  ) =\pm  \sqrt{v^2 + w^2 + 2 v w  \cos\big(k  +\frac{\varphi}{L} \big)+U^2 } \quad. 
\end{equation}
As is well known, the spectrum (\ref{spectrum}) is gapped in the presence of either dimerization ($v \neq w$)  or on-site staggered potential ($U \neq 0$), and gapless otherwise,  while the flux phase $\varphi$ leads to a shift in the momenta. The quantity $\mathsf{v}_m={\rm max} [\partial_k \varepsilon(k)]$ identifying the maximal (dimensionless) group velocity characterizing the excitations along the ring  is given by  
\begin{equation}
\label{vm-def}
\mathsf{v}_m=\sqrt{\frac{\mathsf{A}-\sqrt{\mathsf{A}^2-\mathsf{B}^2}}{2}} \quad,
\end{equation}
where $\mathsf{A}=v^2+w^2+U^2$ and $\mathsf{B}=2 v w$.  The single particle eigenstates $|k, \pm\rangle=|k\rangle \otimes |u_\pm(k)\rangle$ are determined through the eigenvalue problem $(\hat{\mathbf{d}}(k) \cdot \boldsymbol\sigma)  |u_\pm(k)\rangle=\pm  |u_\pm(k)\rangle$, where $\hat{\mathbf{d}}(k)=\mathbf{d}(k)/|\mathbf{d}(k)|$ is the unit vector field. \\
In the SSH model ($U=0$) the two regimes $v>w$ and $v<w$ are known to identify two topologically distinct insulators\cite{ungheresi_book}.

\subsection{Local vs non-local $\mathcal{C}$-symmetry breaking.} 
Similarly to a quantum state $\boldsymbol\rho$, a Hamiltonian ${H}$ is    charge conjugation symmetric if there exists a set $\{ \theta  \}$ of phases such that  $\mathcal{C}^{}_\theta  {H} \mathcal{C}_\theta^\dagger={H}$. On account of the above discussion about gauge 
transformations, one can equivalently say that ${H}$ is   charge conjugation symmetric if there exists a gauge transformation (\ref{gauge-def}) such   that $\mathcal{C}_0^{}  {H}(\{\theta\}) \mathcal{C}_0^\dagger={H}(\{\theta\})$, where ${H}(\{\theta\})$ is the Hamiltonian ${H}$ re-expressed as a function of the gauge operators $\tilde{c}^{}_{j\alpha}\,,\, \tilde{c}^\dagger_{j\alpha}$.

It is straightforward to verify that  the Hamiltonian Eq.(\ref{Ham-real})  {\it does not} exhibit charge conjugation-symmetry, unless $U  = 0$ and $\varphi= m \pi$. 
In particular, one can observe that the $\mathcal{C}$-symmetry is broken {\it locally}  by the on-site potential $U\neq 0$. Indeed, for any site $(j,\alpha)$,  
the projection 
\begin{equation}
{H}_{j\alpha} = \mathcal{P}_{j\alpha} {H} \mathcal{P}_{j\alpha}
\end{equation} 
of the Hamiltonian on that site always  fulfills ${H}_{j\alpha}-\mathcal{C}_\theta {H}_{j\alpha}\mathcal{C}_\theta^\dagger \neq 0$, for all choices of the $\{\theta \}$ phases in Eq.(\ref{CC-def}).
 Equivalently, there is no gauge transformation (\ref{gauge-def}) that 
 can remove the $\mathcal{C}$-breaking  potential $U$ present in that site $(j,\alpha)$.

In contrast,  the presence of the flux phase $\varphi \neq m \pi$ breaks~$\mathcal{C}$ {\it non-locally}, meaning that for  any fixed local portion $P$ of the ring   there is always a   phase set $\{\theta\}$   leading to  ${H}_P-\mathcal{C}_\theta{H}_P\mathcal{C}^\dagger_\theta = 0$, where
${H}_{P} = \mathcal{P}_{P} {H} \mathcal{P}_{P}$ is the Hamiltonian projected on that portion. For instance, by choosing the following  set of linearly growing  phases,
\begin{equation}\label{theta-ell}
\{\theta_\ell \} \equiv \begin{cases}
\theta_{j,A} = \frac{  \varphi}{L} \, (j-\frac{1}{4})  \\
\theta_{j,B} = \frac{ \varphi}{L} \, (j+\frac{1}{4})   \\
\end{cases}  \quad j=1,\ldots L\,\,\,,
\end{equation}
the $\mathcal{C}$-symmetry is realized almost everywhere in the ring, except in the link between sites $(L,B)$ and  $(1,A)$, where $\mathcal{C}$-symmetry is broken.\cite{nota-translational-symmetry} In other words, the
gauge transformation (\ref{gauge-def}) with phases (\ref{theta-ell})   accumulates the entire flux phase in such link. Similar phase choices accumulate the $\mathcal{C}$-breaking elsewhere. Note that the possibility of performing
the gauge transformation is the reason why the model Eq.(\ref{Ham-real}), when defined  in a lattice with {\it open} boundary conditions (OBCs), i.e. on a chain,  is $\mathcal{C}$-symmetric. Indeed, since a chain can be regarded  as a ring with one missing  link,  it is always possible to get rid of the phase $\varphi$ appearing in the tunneling amplitudes by ``accumulating" it  along such a link. In contrast, in the ring-shaped geometry considered here, the PBCs imply that  ${H}-\mathcal{C}_\theta^{} {H}\mathcal{C}_\theta^\dagger = 0$ cannot be realized in all links, and at least one $\mathcal{C}$-breaking link  will always be present for topological reasons.

\section{tsunami effect}
\label{sec4}
\subsection{$\mathcal{C}$-symmetry breaking   for the ground state}
We start from some preliminary analysis  at equilibrium.
When the Hamiltonian (\ref{Ham-real}) fulfills the $\mathcal{C}$-symmetry ($U=0$ and $\varphi =\pi m$),   its non-degenerate ground state at half filling also does. 
If the $\mathcal{C}$-symmetry of the Hamiltonian is   explicitly broken by $U \neq 0$ or  $\varphi \neq \pi m$, this will reflect on its ground state. In particular, because of the dimerization, one expects the density  to  deviate from Eq.(\ref{TH2}) and to acquire the form
\begin{equation}\label{deviation}
{n}_{A/B}= \langle {n}_{j, A/B} \rangle = \frac{1}{2} \pm \delta n  \hspace{1cm} \forall j\quad,
\end{equation}
where the deviation  $\delta n$, which represents the sublattice unbalance, is independent of the cell $j$ because of   the cell translational invariance of model (\ref{Ham-real}).

However, for $U=0$, i.e. in the SSH model, Eq.(\ref{TH2}) still holds strictly  for any length $L$, despite the $\mathcal{C}$-breaking   flux $\varphi$. The reason is that such a model exhibits the additional   chiral symmetry $\mathcal{S}$,\cite{ungheresi_book} which forbids  deviations from Eq.(\ref{TH2}) for the ground state, even in the presence of the flux.\cite{NJP-rossi-dolcini} This can be explicitly seen from the general expression 
\begin{equation}
\delta n =-\frac{1}{2L} \sum_{k=-\pi}^{\pi} \hat{d}_z(k)\quad,
\end{equation}
where $\hat{d}_z(k)$ is the $z$-component of the unit vector related to $\mathbf{d}(k)$ in Eq.(\ref{d-vec}), and  vanishes in the SSH model. In contrast, in the RM model, the staggered on-site potential $U\neq 0$ does lead to a deviation $\delta n$ in Eq.(\ref{deviation}), which
 in the thermodynamic limit reads
  \begin{equation}
  \delta n=- \frac{1}{\pi}\frac{U}{\sqrt{U^2+(v+w)^2}}\, \mathsf{K}\left(\frac{4 v w}{U^2+(v+w)^2}\right)\quad,
  \end{equation}
where $\mathsf{K}$ is the complete elliptic integral of the first kind.

\subsection{Quench dynamics}
\label{sec4B}
While at equilibrium   the effects of $\mathcal{C}$-breaking induced by the flux may be masked  by the additional chiral symmetry $\mathcal{S}$, when the system is driven out of equilibrium such a symmetry gets broken  {\it dynamically} because of its anti-unitary character\cite{Cooper_PRL_2018,Cooper_PRB_2019}, and its protection is lost in the dynamically evolving quantum state. At first, one can thus expect   deviations from Eq.(\ref{TH2})    to emerge also in the SSH model right after the quench. As we shall see, this is not the case, though.
Let us thus analyze a quantum quench protocol, where the system is initially prepared in the ground  state $\boldsymbol\rho^i$  of some initial (pre-quench) Hamiltonian ${H}^i$. At $t=0$  the Hamiltonian parameters are suddenly changed   (${H}^i\rightarrow {H}^f$), so that the   dynamical evolution $\boldsymbol\rho(t)=\exp[-i {H}^f t/\hbar] \boldsymbol\rho^i\exp[+i {H}^f t/\hbar]$ is governed by the  post-quench Hamiltonian ${H}^f$.
We denote by $\mathbf{d}^i(k)$ and $\mathbf{d}^f(k)$ the vector fields characterizing  ${H}^i$ and ${H}^f$, respectively, in Eq.(\ref{Ham-k}). As motivated in the Introduction, the   $\mathcal{C}$-breaking parameters, namely the flux phase $\varphi$ and the local on-site potential $U$, will be kept constant across the quench.  We shall mainly focus here on   the quench protocol that  exchanges the hopping amplitudes, i.e. 
\begin{equation}
\begin{array}{lcccl}
v^i =1 & & \rightarrow  & &v^f= r  \\
w^i= r & & \rightarrow & & w^f=1
\end{array}\quad, \label{quench}
\end{equation}
where  $r$  is a dimensionless parameter characterizing the dimerization strength ($0<r<1$). The cases of other quench protocols will be addressed later. 
For the sake of completeness, before presenting the results, a few technical details are in order. The initial single particle density matrix describing the ground state of ${H}^i$ with its  fully occupied lower band  is block-diagonal in $k$-space, $\rho^i=\oplus_k \rho^i(k)$, where $\rho^i(k)=|u^i_{-}(k)\rangle \langle u^i_{-}(k)|=(\sigma_0-\hat{\mathbf{d}}^i(k)\cdot \boldsymbol\sigma)/2$,   $\sigma_0$ denotes the $2 \times 2$ identity matrix, and $\hat{\mathbf{d}}^i(k)=\mathbf{d}^i(k)/|\mathbf{d}^i(k)|$  is the initial unit vector field. 
Because the quench dynamics is decoupled is $k$-space, the $k$-block of the evolved density matrix  is $\rho_{-}(k,t)=|u_{-}(k,t)\rangle \langle u_{-}(k,t)| $, where $|u_{-}(k,t)\rangle =\exp[-i \mathbf{d}^f(k) \cdot \boldsymbol\sigma \gamma t/\hbar] |u_{-}^i(k)\rangle$, and can always be written as  
$\rho_{-}(k,t)=   [ \sigma_0 - \hat{\mathbf{d}}(k,t)\cdot \boldsymbol{\sigma}  ]/2$, where   $\hat{\mathbf{d}}(k,t)$ is  a time-dependent unit vector  given by \cite{chen_PRB_2018}
\begin{eqnarray}
 \hat{\mathbf{d}}(k,t)&=& \mathbf{d}_\parallel(k)+\mathbf{d}_\bot(k) \cos[2|\mathbf{d}^f(k)|\tau]+ \nonumber \\
 & & +\mathbf{d}_\times(k)\sin[2|\mathbf{d}^f(k)|\tau] \quad, \label{hatd(k,t)}
 \end{eqnarray}
where $\mathbf{d}_\parallel(k)= [ \hat{\mathbf{d}}^i(k) \cdot \hat{\mathbf{d}}^f(k) ]\hat{\mathbf{d}}^f(k)$,   $\mathbf{d}_\bot(k) = \hat{\mathbf{d}}^i(k) - \mathbf{d}_\parallel(k)$ and $\mathbf{d}_\times(k) = -[ \hat{\mathbf{d}}^i(k) \times \hat{\mathbf{d}}^f(k) ]$, with $\hat{\mathbf{d}}^f(k)=\mathbf{d}^f(k)/|\mathbf{d}^f(k)|$, and
\begin{equation}\label{tau-def}
\tau = \frac{t \, \gamma}{\hbar } 
\end{equation}
denotes the dimensionless time.\\

The knowledge of the single-particle density matrix~$\rho$, when rewritten in real space through $\rho_{j_1 \alpha_1,j_2 \alpha_2}(t)=L^{-1} \sum_k e^{i k (j_1-j_2)} (\rho_{-})_{\alpha_1 \alpha_2}(k,t)$  [see Eq.(\ref{SP-rho-def})], straightforwardly provides the dynamical evolution of the  site density expectation value (diagonal entries) and all two-point correlations (off-diagonal entries). Here below we shall   illustrate the dynamical behavior of these quantities.
\begin{figure}[h]
\centering
\includegraphics[width=  \columnwidth]{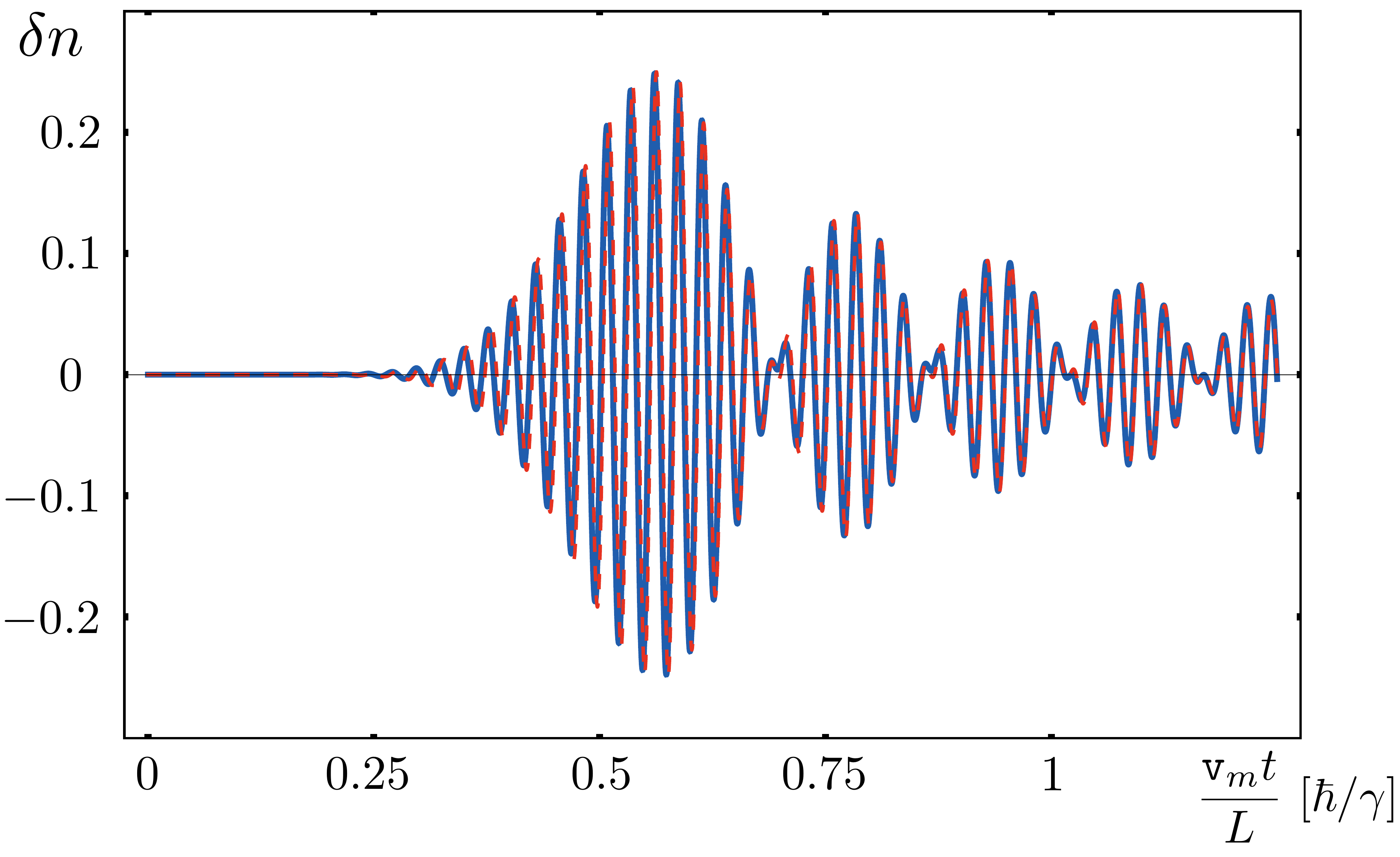}
\caption{\label{Fig2}    Time evolution of the on-site density deviation [see Eq.(\ref{deviation})]   in a  SSH ring with $L=12$ cells (=24 sites), threaded by a magnetic flux $\varphi=\pi/2$, undergoing 
the quench  protocol (\ref{quench}), with dimerization $r=0.1$. The analytical expression (\ref{deltanA-asympt}) obtained in the  strongly dimerized limit (red dashed curve) perfectly reproduces the numerically exact evolution (blue curve). The deviation $\delta n$ is exponentially small in the system size $L$ [see Eq.(\ref{expo})] for an extensively long time. }
\end{figure}

\subsubsection{On-site density and currents}
{\it Density.} Even  far from equilibrium, the cell translational invariance enables us to write  the on-site density as $n_{A/B}(t)=1/2\, \pm \delta n(t)$, and it is thus sufficient to analyze the dynamical behavior of the deviation $\delta n(t)$, which is independent of the cell $j$. 
We recall that,   if the $\mathcal{C}$-symmetry were  fulfilled by the initial state and by the Hamiltonian, i.e. if $U=0$ and $\varphi=0$, the density would be pinned to $1/2$  [see  Eq.(\ref{pinning})] and the deviation would be strictly vanishing at any time, $\delta n(t) \equiv 0$. In the case of the RM model  without flux ($\varphi=0$), where the $\mathcal{C}$-symmetry breaking is induced locally by the on-site staggered potential $U \neq 0$, the deviation is already non-vanishing in the pre-quench state, and after the quench it significantly fluctuates in time, as displayed in Fig.\ref{Fig1}(a). In contrast, in the SSH model ($U=0$), where $\mathcal{C}$-symmetry is broken   non-locally by the presence of a flux  $\varphi $, the density   remains effectively pinned to $1/2$ for a strikingly long time, proportional to the system size~$L$, after which the deviation starts  to appreciably fluctuate [see Fig.\ref{Fig1}(b)]. This is the tsunami effect arising from quantum dynamics.   In order to gain more quantitative information, we have investigated the limit of a  strongly dimerized ring ($r  \ll 1$), where it is possible to find an analytical expression for the density behavior in the considered quench protocol (\ref{quench}). In particular, for even $L$ one has 
\begin{equation}\label{deltanA-asympt}
\delta n(\tau) \simeq \frac{  L}{2\tau\,r} (-1)^{\frac{L+2}{2}}  \, J_L(2 \tau r) \, \cos(2\tau) \sin \varphi\quad,
\end{equation}
with $J_L$ denoting the Bessel function of order $L$. A similar expression can be obtained for odd $L$ [see App.\ref{AppB}]. Equation~(\ref{deltanA-asympt}) holds for times $2 \tau r < 2 L$, as can be seen from Fig.\ref{Fig2}, where we have compared the exact numerical evolution (solid blue curve)  with the asymptotic approximation (dashed red curve), for a dimerization $r=0.1$. 
While technical details of such derivation can be found in Appendix~\ref{AppB}, here we emphasize the insights gained from Eq.(\ref{deltanA-asympt}). First, the flux determines the maximal magnitude of density deviation, which vanish  in the $\mathcal{C}$-symmetric case ($\varphi=\pi m$), as expected, and are maximal for $\varphi=\pi (m+1/2)$ with $m \in \mathbb{Z}$. Second, the asymptotic expansion of the Bessel function shows that for the time range $2\tau r\ll \sqrt{L}$  one has
\begin{equation}\label{expo}
\delta n (\tau) \simeq \left( \frac{\tau r\, e}{L}\right)^L \quad,
\end{equation}
i.e. the deviations from $1/2$ are suppressed {\it exponentially} with the system size $L$, whereas for longer time $L \lesssim 2 \tau r < 2 L$ the deviation $\delta n (\tau)$ acquires a double period structure, namely  a longer period envelope function    $J_L(2 \tau r)\sim \cos(2 \tau r-\pi/4)/\sqrt{\pi \tau r}$, characterized by an  {\it algebraic}  decay,  multiplied  by the shorter period oscillatory  function  $\cos(2\tau)$. Finally, by noticing that in the SSH model the maximal velocity (\ref{vm-def}) reduces to $\mathsf{v}_m=w=r$, the time
\begin{equation}\label{tau-2-star}
\tau^*_2=\frac{L}{2r}= \frac{L}{2 \mathsf{v}_m}
\end{equation}
represents an estimate of the onset of the tsunami effect, i.e. 
the time characterizing  a dynamical crossover  between the exponential and the algebraic suppression in~$L$ of the density deviation $\delta n$. Roughly, the onset time $\tau_2^*$   also represents the time,  at which   $\delta n$ reaches   its first maximal value.  As one can see from Eq.(\ref{tau-2-star}),     $\tau^*_2$ grows linearly  with the system size $L$ while, at fixed~$L$, it  increases for lower values $r$, i.e. for stronger dimerization.

 We emphasize that, while  for the specific protocol (\ref{quench}) it is possible to gain the analytical expression Eq.(\ref{deltanA-asympt}) for the density dynamics, the occurrence of the tsunami effect is by no means restricted to such case. A qualitatively very similar result emerges both if the quench is performed across  the  two topologically distinct phases and within the same topological phase, and also when hopping terms between next-to-nearest neighbor cells are included. While the magnitude of the density deviation $\delta n$ after the tsunami onset quantitatively depends on the specific values of the pre-quench and post-quench parameters,   the tsunami effect is a quite generic phenomenon occurring for any  quench, provided that the $\mathcal{C}$-symmetry is broken non-locally by the flux and   the initial state is insulating ($v^i \neq w^i$). \\
 
The importance of an initially insulating state for the tsunami effect can be understood by comparing the dynamical density behavior in the case where the initial state is the half-filled ground state of the non-dimerized Hamiltonian ${H}^i$ with flux, i.e. a metallic ground state. By performing a quench to a dimerized ${H}^f$ through the protocol   
\begin{equation}
\begin{array}{lcccl}
v^i =1 & & \rightarrow  & &v^f= 1  \\
w^i= 1 & & \rightarrow & & w^f=r
\end{array}\quad, \label{quench-metal}
\end{equation}
one obtains   a density deviation emerging immediately after the quench, and scaling as $O(1/L)$. Indeed already  for  short times $\tau \ll L/\mathsf{v}_m$ and small flux $0<\varphi \ll \pi/2$  one can find for odd $L$  
\begin{equation}\label{deltan-metal}
 \delta n (\tau)\simeq -\frac{\sin [2  (1-r) \tau]}{\pi}   \frac{\varphi}{2L}
\end{equation}
indicating that no tsunami effect is present for an initially metallic state. A similar formula is found for even $L$.\\

{\it Currents.} When the tsunami effect starts to manifest itself in the SSH model (see Fig.\ref{Fig2}), the densities in sites $A$ and $B$ of each cell start to fluctuate oppositely, $n_{A/B}(t)=1/2\, \pm \delta n(t)$. These fluctuations represent oscillating dipoles, which in turn generate non-stationary currents. Specifically, in a dimerized lattice one has   inter-cell and intra-cell particle  currents, given by
\begin{eqnarray}\label{inter-cell-cur-def}
  \hat{J}_j^{inter}   
&=& \frac{\gamma}{\hbar} w     \left( i e^{i \frac{\varphi}{2L}}  {c}_{jB}^\dagger  {c}_{j+1A}- i e^{-i \frac{\varphi}{2L}}  {c}_{j+1A}^\dagger  {c}_{jB}  \right) 
\end{eqnarray}
and
\begin{eqnarray}\label{intra-cell-cur-def}
    \hat{J}_j^{intra}   =       \frac{\gamma}{\hbar} v \left(  i e^{i \frac{\varphi}{2L}}  {c}_{jA}^\dagger {c}_{jB}- i e^{-i \frac{\varphi}{2L}}   {c}_{jB}^\dagger {c}_{jA}\right)  \quad, 
    \end{eqnarray}
respectively. Their expectation values can be computed by exploiting the cell translational invariance
\begin{equation}
J^{\nu}   =\frac{1}{L} \sum_{j=1}^L \langle \hat{J}_j^{\nu}\rangle \hspace{1cm} \nu=inter/intra
\end{equation}
and   by means of the single-particle density matrix (\ref{SP-rho-def}). From  the continuity equation $\partial_t (\delta n)=     J^{inter}- J^{intra}$, one can   straightforwardly deduce that the tsunami effect characterizing $\delta n$ is also present in the currents (plots not shown here).

\subsubsection{Correlations}
The off-diagonal entries of the single-particle density matrix  (\ref{SP-rho-def}) describe the two-point equal time correlation functions, which only depend on the  relative distance $l$ between two cells, due to cell translational invariance.  For definiteness, we  have taken~$L$   to be even and,  focussing on the ``central" cell $j^*=L/2$ as a reference cell, we have analyzed its correlation
 $\rho_{L/2+l,\alpha;  L/2,\beta}(t)$ with any other   cell located at an arbitrary distance   $l\in [-L/2+1,L/2]$  from $j^*$, with $\alpha,\beta=A,B$. We want to investigate    when   the presence of the magnetic flux manifests itself in correlations.
\begin{figure}[h]
\centering
\includegraphics[width=7 cm]{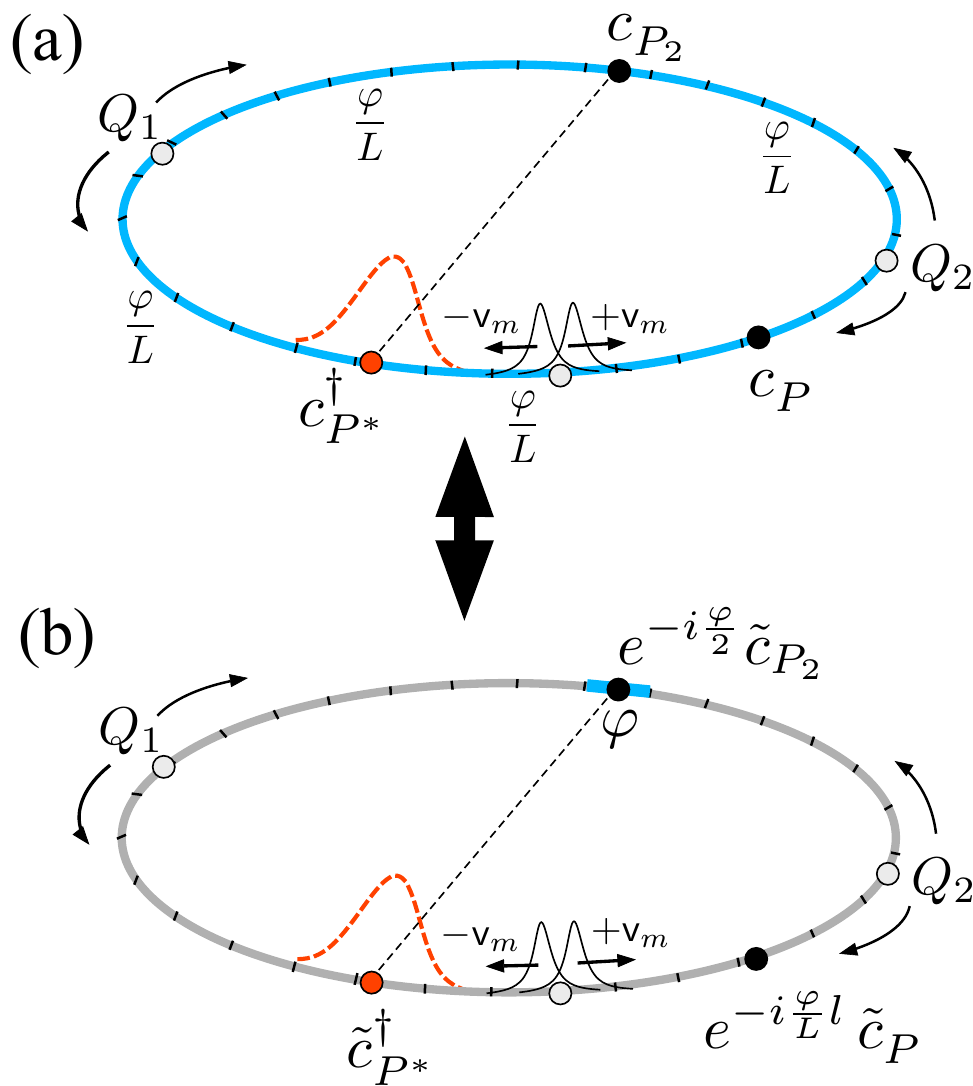}
\caption{\label{Fig3} Correlations between the reference point $P^*$ (red dot) and any other point $P$ along the ring (black dots) evolve in time, as a result of the quench
because pairs of counter-propagating excitations depart from   intermediate ring points (grey dots) and reach  the envisaged red and black points.   The non-local presence of the flux can be observed only when two pairs of excitations, departed from intermediate  points  $Q_1$ and $Q_2$, have covered   the {\it entire} ring and have reached the  two points   (red and black) simultaneously.
The two panels illustrate
two types of gauges   describing the presence of the magnetic flux threading the ring. (a) the flux phase $\varphi$ is uniformly distributed along the ring, so that the vector potential along each link is $O(\varphi/L)$. (b) the flux phase $\varphi$ is accumulated only along the link located oppositely to the reference point $P^*$.  }
\end{figure}

Let $P^*=(j^*,A)$ denote a reference  point, namely the $A$-site of the reference cell,  and $P=(j^*+l,A)$ another $A$-site, as depicted  by the red and black dots in Fig.\ref{Fig3}(a), respectively. As a result of the quench, excitations emerging from the midpoint (grey dot) between $P^*$ and $P$ start to propagate in opposite directions, with typical group velocities $\pm \mathsf{v}_m$ of the post-quench Hamiltonian ${H}^f$ [see Eq.(\ref{vm-def})], and    reach  at the same time the two points $P^*$ and $P$, which thus get correlated. We emphasize that, although these excitations travelling along the ring arc see the local phase $\varphi/L$ across each cell (depicted in blue in Fig.\ref{Fig3}), this cannot encode the non-local effect of the flux.  Indeed,  at such time these excitations have only travelled a {\it portion} of the ring and cannot distinguish a  chain with OBCs, which is $\mathcal{C}$-symmetric, from  a ring with PBCs threaded by flux that breaks
$\mathcal{C}$-symmetry. The   presence of the flux can manifest itself only  after excitations have explored the {\it entire} ring. 
 At a more formal level, this can be seen by applying the gauge transformation  (\ref{gauge-def}) with Eq.(\ref{theta-ell}), which removes   the   tunneling amplitude complex phases from almost all links in the Hamiltonian (\ref{Ham-real}) and accumulates  the entire flux phase $\varphi$ along the link opposite to the reference point $P^*$, as shown by the blue link at $P_2$ in Fig.\ref{Fig3}(b). In this way the correlation function between (say) $A$ sites is rewritten as
\begin{equation}\label{corr-single-out}
\rho_{\frac{L}{2}+l,A; \frac{L}{2}, A}=  \langle c^{\dagger}_{\frac{L}{2},A} c^{}_{\frac{L}{2}+l,A}\rangle =e^{-i \frac{\varphi}{L} l}  \tilde{\rho}_{\frac{L}{2}+lA,\frac{L}{2}A} \quad,
\end{equation}
where $e^{-i \frac{\varphi}{L} l}$ is a trivial time-independent phase factor, while $\tilde{\rho}_{L/2+lA; L/2,A}=\langle \tilde{c}^{\dagger}_{L/2,A} \tilde{c}^{}_{L/2+l,A}\rangle $ is the correlation in the new gauge.

\begin{figure}[h]
\centering
\includegraphics[width=\columnwidth]{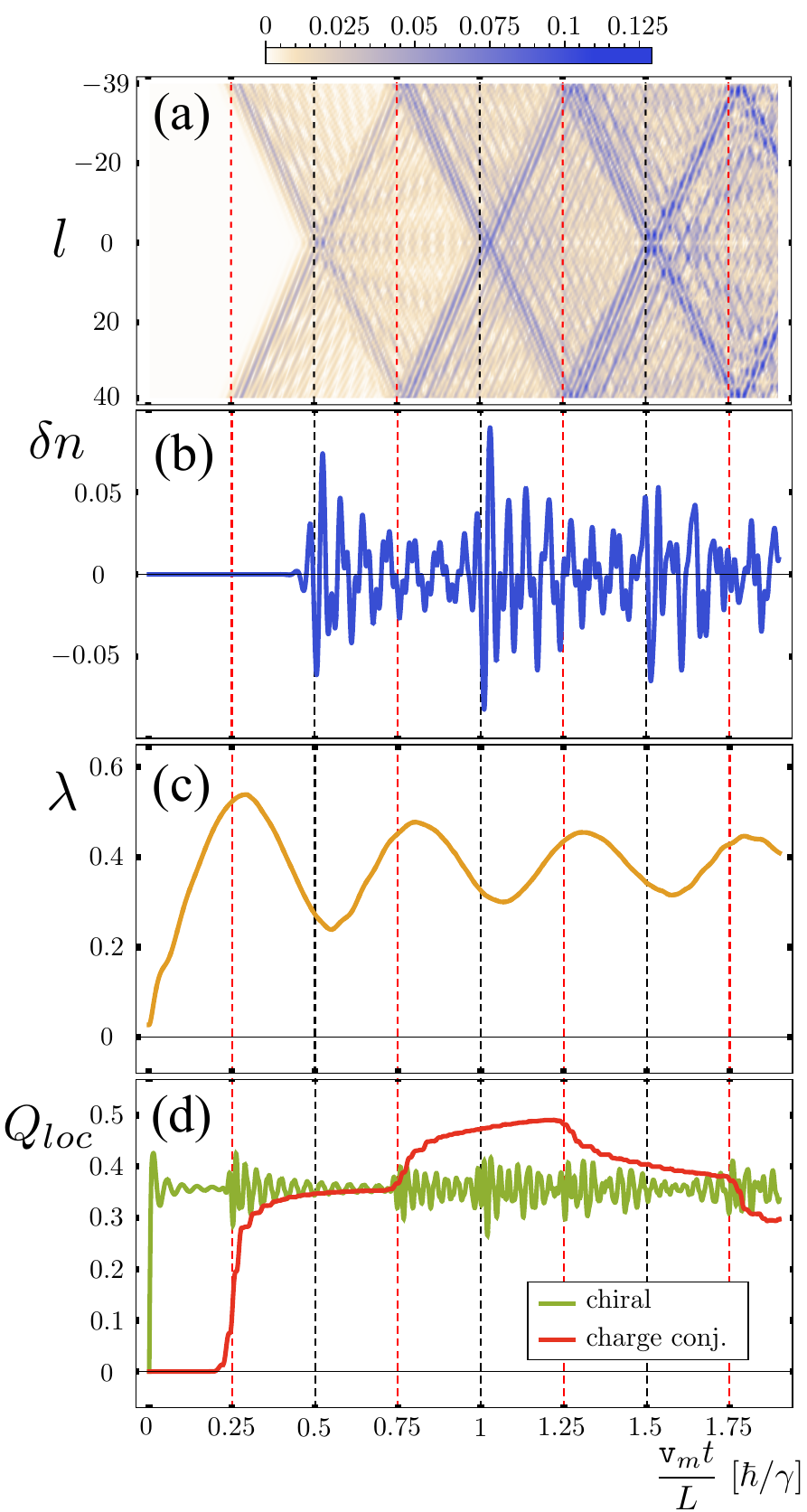}
\caption{\label{Fig4} Dynamics of the SSH ring [Eq.(\ref{Ham-real}) with $U=0$], with $L=80$ cells ($=160$ sites), threaded by a flux ($\varphi=\pi/4$), quenched with the protocol (\ref{quench}), with dimerization parameter $r=0.7$. (a) density plot of the space-time behavior of correlations, with $l\in [-L/2,L/2]$ denoting the distance between two sites. (b) The dynamical behavior of the site density deviation $\delta n$ from $1/2$ [see Eq.(\ref{deviation})]. (c) The time evolution of the maximally localized Wannier functions. (d) Time-evolution of the quantifiers of local  breaking for chiral symmetry (green curve) and charge-conjugation symmetry (red curve).}
\end{figure}

Such a new gauge enables us to highlight the crucial role of the localization properties of the quantum state. Indeed, if the correlation term  $\tilde{\rho}_{L/2+lA; L/2,A}$ in Eq.(\ref{corr-single-out}) has a quasi long range, i.e. a spatially slow power law decay in $l$, as is the case in a metal, the reference point $P^*$ does feel the presence of the flux accumulated at the opposite site $P_2$ of the ring, even in the initial pre-quench state. In contrast, if the state is characterized by exponentially decaying correlations, as is the case in an insulator, $\tilde{\rho}_{L/2+lA; L/2,A}$  is  initially independent of $\varphi$, and the flux phase in $P_2$ remains   invisible to $P^*$.
   In order to analyze when a non-trivial flux dependence emerges in an insulator, we have thus first singled out the trivial phase factor in Eq.(\ref{corr-single-out}) and then subtracted the correlations in the absence of flux. Explicitly, the quantity
\begin{equation}\label{Delta-rho}
\Delta \rho_{\frac{L}{2}+l, \alpha_2; \frac{L}{2},  \alpha_1} \doteq \left. \tilde{\rho}_{\frac{L}{2}+l ,\alpha_2; \frac{L}{2} , \alpha_1}\right|_{\varphi\neq 0}\, -  \left.\rho_{\frac{L}{2}+l, \alpha_2; \frac{L}{2},  \alpha_1}  \right|_{\varphi=0} 
\end{equation}
encodes the non-trivial dynamical effects of the flux on the quenched system.
In Fig.\ref{Fig4}(a) we have plotted $|\Delta \rho|$ of Eq.(\ref{Delta-rho}) for $\alpha_1=\alpha_2=A$ as a function of time and cell distance $l$, in  a quenched SSH model of $L=80$ cells ($=160$ sites), with dimerization parameter $r=0.7$, and flux phase $\varphi=\pi/4$. The  plot shows that the tsunami effect is present also in the correlations, since the difference  $\Delta \rho$ from the zero-flux case is vanishing until the extensive time 
\begin{equation}\label{tau-1-star}
\tau^*_1=  \frac{L}{4 \mathsf{v}_m}\quad,
\end{equation}
when  a $\Delta \rho\neq 0$  starts to be visible  at the maximal distance $l=\pm L/2=\pm 40$. Only after $\tau_1^*$ a non-trivial flux dependence arises between the red reference point 
 $P^*$ and the black point $P_2$   located symmetrically with respect to it (see  Fig.\ref{Fig3}). The time (\ref{tau-1-star}), highlighted in Fig.\ref{Fig4} by a vertical red dashed line, can be interpreted
as the time when two pairs of counter-propagating excitations  departing after the quench from the points $Q_1$ and $Q_2$ (grey dots) simultaneously reach $P^*$ and $P_2$, after travelling a distance $L/4$ each, thereby covering the entire ring and probing the presence of the flux. This time is of course independent of the chosen gauge (a) or (b) in   Fig.\ref{Fig3}.

After the time $\tau_1^*$, the correlation  $\Delta \rho$ dynamically propagates    along the ring following two symmetrical light-cone trajectories in space time [blue lines in Fig.\ref{Fig4}(a)], characterized by a velocity $2\mathsf{v}_m$  resulting from the pair of counter-propagating excitations travelling at $\pm \mathsf{v}_m$.
 These light cones intersect at distance $l=0$ at the time~$\tau_2^*$ given by Eq.(\ref{tau-2-star}), as highlighted by a vertical black dashed line. Notice that  the $A$-$A$ correlation difference $\Delta \rho$ at ``zero-distance" ($l=0$) is nothing but the on-site density deviation   $\delta n=\langle {n}_{L/2,A} \rangle-\langle {n}_{L/2,A} \rangle|_{\varphi=0} $. This is shown in Fig.\ref{Fig4}(b), which represents a cut at $l=0$ of the correlation plot displayed in panel (a).
Notably, the time $\tau_1^*$ in Eq.(\ref{tau-1-star}) is a half of the time $\tau_2^*$ in Eq.(\ref{tau-2-star}). This is because a non-local effect such as the presence of the flux impacts on non-local correlations {\it earlier} than the local density, which thus experiences the tsunami effect onset as   last. Indeed, referring to  Fig.\ref{Fig3}, the density in $P^*$ can be considered as the correlation between $P^*$ and $P \equiv P^*$, and the effect of the flux can only appear when a pair of counter-propagating excitations departing from the point $P_2$ have travelled a distance $L/2$ each, thereby probing the flux presence over the entire ring.

Correlation  lightcones then continue to  evolve   in the ring space-time as shown in Fig.\ref{Fig4}(a),  where the bottom and top are identified simply because of the PBCs. Notice that   crossings at minimal distance $l=0$  always occur at times $\tau^*_{2n}=2n \tau_1^*=n  \tau_2^*$, as highlighted by the black vertical dashed lines, and roughly correspond to the maximal peaks of the density shown in Fig.\ref{Fig4}(b). Similarly, correlations at maximal distance $l=\pm L/2$ always occur at times $\tau^*_{2n+1}=(2n+1) \tau_1^*=(n+1/2)\tau_2^*$,  as highlighted by the red vertical dashed lines. At very long times, one observes that light cones start to become less sharp, due to the  dispersion  related to the band curvature.\\  

\subsubsection{Wannier Localization length}
We have thus seen that, when the $\mathcal{C}$-symmetry is broken, the dynamical response to a quench dramatically depends on whether the symmetry is broken locally, like in the RM model ($U \neq 0$), or non-locally, like in the SSH model with flux ($\varphi \neq 0$). In particular, in the latter case the local observables and correlations appear  effectively robust to the $\mathcal{C}$-breaking and behave as if $\mathcal{C}$-symmetry were preserved for an extensively long time, as shown in Fig.\ref{Fig1} and \ref{Fig4}.
Here we would like to interpret this effect in terms of Wannier functions. Indeed we recall that  the ground state of a band insulator    can always be considered as a Slater determinant of Wannier functions, localized around the various lattice cells.
In particular it is possible to identify maximally localized Wannier functions (MLWFs), whose spread  provides the physical localization length of the system.
While in the thermodynamic limit ($L\rightarrow \infty$), the definition of MLWFs is well established\cite{kohn_PR_1959,vanderbild_PRB_1997,vanderbild_PRB_2002,vanderbild_RMP_2012},   in the case of finite size systems   with PBCs analyzed here,  their derivation   requires some care. However, this can be done by harnessing concepts of directional statistics\cite{fisher-book}, and an angular localization length~$\lambda$ can be identified  in the ring (details of this derivation  are given   in Appendix~\ref{AppC}).
 This quantity   enables us to understand the difference between the local {\it vs} non-local $\mathcal{C}$-symmetry breaking. Indeed  in the RM insulator a localized function is directly affected by a local potential $U\neq 0$ present within its localization length. In contrast, in the SSH insulator a Wannier function cannot be   sensibly affected by the flux $\varphi$, which requires  to probe the entire ring. This is clearly seen by invoking again the gauge transformation illustrated in Fig.\ref{Fig3}(b), where the Wannier function centered around the reference point $P$ (sketched as a red dashed Gaussian curve) cannot ``see" the flux accumulated on the other side of the ring.
In contrast, in a metal, where Wannier functions are delocalized, namely algebraically decaying in space\cite{kohn_PR_1959,vanderbild_PRB_1997,vanderbild_PRB_2002,vanderbild_RMP_2012}, the non-local $\mathcal{C}$-symmetry breaking due to the flux can be felt by the ground state.

Focussing now on the SSH model, we have also 
 determined the dynamical evolution of the MLWF localization length $\lambda$ after the quench.
The result is displayed in Fig.\ref{Fig4}(c) and shows that $\lambda$ starts to increase after the quench. We recall  that the Wannier (angular) localization length $\lambda$  also determines the typical lengthscale, over which two ring cells are correlated. Thus, the  lightcones shown in Fig.\ref{Fig4}(a) and describing the  evolution of correlations in space-time can also be seen as the dynamical spreading of the  Wannier localization length $\lambda$.

The increase of $\lambda$ reaches a maximal value roughly at the   time~$\tau_1^*$. By comparing with Fig.\ref{Fig4}(a), we recall that~$\tau_1^*$ is the time when the tsunami effect appears for the very first time, and it occurs in the correlation length of two maximally separated points ($l=\pm L/2$), as highlighted by the red dashed vertical line in Fig.\ref{Fig4}. This means that~$\tau_1^*$ corresponds to the time where the Wannier functions have widespread enough to explore  the entire ring and to probe the existence of the flux. 
A closer inspection shows that the maximum of the Wannier spreading actually occurs with a slight delay with respect to~$\tau_1^*$. This is due to the fact that, because of the curvature of the band dispersion relation, excitations do not all propagate at the same velocity, as is also clear from the color fringes in in Fig.\ref{Fig4}(a). Thus, while~$\tau_1^*$ corresponds to the earliest arrival of the counter-propagating excitations travelling with maximal velocity $\mathsf{v}_m$, the maximal spreading of the Wannier function occurs when also slowlier excitations have travelled the distance $L/2$. 

After reaching the maximum, the localization length $\lambda$ decreases and reaches a local minimum roughly at time $\tau_2^*$, which is the latest time  when the tsunami effect manifests itself, and it occurs at the minimal distance correlation ($l=0$), namely the on-site density [see Fig.\ref{Fig4}(b),  and black dashed vertical lines]. 
After such time, $\lambda(t)$ exhibits an oscillatory behavior, where  minima and maxima roughly occur at the times $\tau^*_{2n}$ and  $\tau^*_{2n+1}$, respectively, up to the delays originating from the band curvature.  \\

\subsubsection{Charge conjugation {\it vs} chiral symmetry breaking}
Before concluding this section, a  remark is in order about the different impact of charge conjugation and chiral symmetries at local level.
We recall that the chiral transformation $\mathcal{S}$ is an anti-unitary transformation ($\mathcal{S}^{}  i\mathcal{S}^\dagger =- i$ and $\mathcal{S}^\dagger=\mathcal{S}^{-1}$), whose action 
\begin{equation}
\left\{ \begin{array}{lcl}
\mathcal{S} {c}^\dagger_{jA}\mathcal{S}^\dagger&=&+{c}^{}_{jA} \\ & & \\ \mathcal{S} {c}^\dagger_{jB}\mathcal{S}^\dagger &=& -{c}^{}_{jB}
\end{array}
\right. \label{S-def-pre}
\end{equation}
can be compactly written as
\begin{equation}\label{S-def}
    \mathcal{S}^{}  {c}^{\dagger}_{j \alpha} \mathcal{S}^\dagger =  (\mathsf{U}_0)_{j\alpha; l \beta} \, {c}^{}_{l \beta} \quad,
\end{equation}
where $\mathsf{U}_0$ is given in Eq.(\ref{U0-def}). 
Notice that, differently from $\mathcal{C}$,  the action of $\mathcal{S}$ is unaffected by a gauge transformation (\ref{gauge-def}), due to its  anti-linear character.

The SSH Hamiltonian ${H}_{\rm SSH}$  [i.e. Eq.(\ref{Ham-real}) with $U=0$] is known to fulfill the chiral symmetry\cite{ungheresi_book} ($ \mathcal{S}^{} {H}_{\rm SSH} \mathcal{S}^\dagger= {H}_{\rm SSH} $) and so does its ground state, even in the presence of a magnetic flux. Thus, at equilibrium the 
presence of the chiral symmetry $\mathcal{S}$ forces the local density  to   equal $1/2$ in each site, thereby ``masking"
the explicit $\mathcal{C}$-symmetry breaking  due to the flux. However, when the SSH model is driven out of equilibrium by a quench, the $\mathcal{S}$-symmetry is dynamically broken\cite{Cooper_PRL_2018,Cooper_PRB_2019}. This is because, 
despite its similarity with the unitary charge-conjugation $\mathcal{C}_0$ [see Eq.(\ref{C0-def})], $\mathcal{S}$ is   anti-unitary. Physically, this  
  means that the tsunami effect, i.e. the extensively long lasting rigidity to the non-local $\mathcal{C}$-breaking caused by a magnetic flux, is by no means due to the  chiral symmetry protection, which only occurs ``accidentally"   in the equilibrium ground state. 
 
In order to support this conclusion, one can introduce a local quantifier $Q^\nu_{loc}$  of symmetry breaking, which measures at any time ``how much" the symmetry ($\nu=\mathcal{C},\mathcal{S}$) is  broken at local level in the instantaneous   quantum state. While details about the  definition and the evaluation of these quantifiers will be given thoroughly  in the next section and in Appendix~\ref{AppD},  we feel appropriate to anticipate here the dynamical behavior of these local quantifiers $Q^\nu_{loc}$,   illustrated in Fig.\ref{Fig4}(d). Specifically, the green curve depicts the local quantifier of the   chiral symmetry breaking,   which becomes non vanishing immediately after the quench, as a hallmark of the dynamical symmetry breaking of    $\mathcal{S}$. In contrast, the red curve describes the quantifier for   the charge conjugation symmetry  $\mathcal{C}$, which remains exponentially small (in system size) until  the time $\tau_1^*$, i.e. the time at which correlations start to experience the presence of the flux. Only for $\tau> \tau_1^*$ can the SSH model  feel the actual breaking of the $\mathcal{C}$-symmetry. In turn, this also shows that, despite the abrupt dynamical breaking of the chiral symmetry, local observables may still be rigid to dynamical changes.

\section{Symmetry breaking quantifiers}
\label{sec5}
In this section we  introduce  and discuss in details the  quantifiers of  symmetry breaking  that have been anticipated at the end of the previous section. These quantities are  meant to identify ``how much" a given symmetry is broken in a quantum state.   
This will enable us to  quantitatively distinguish the impact of local {\it vs} non-local $\mathcal{C}$-symmetry breaking and also to characterize the difference between  breaking  the discrete symmetries relevant for our problem, namely  charge conjugation symmetry $\mathcal{C}$ and chiral symmetry $\mathcal{S}$.  Similar ideas, based on entanglement, have been recently applied in Ref.[\onlinecite{calabrese_2023}] to  quantify the breaking  of continuous symmetries.

\subsection{General definition of $\mathcal{C}$-symmetry and $\mathcal{S}$-symmetry breaking quantifiers}
\label{sec5a}
\subsubsection{Charge conjugation} We start from the charge conjugation symmetry. As discussed in Sec.\ref{sec2}, if a quantum many-particle state $\boldsymbol\rho$ is $\mathcal{C}$-symmetric ($\mathcal{C}_\theta^{} \boldsymbol\rho\mathcal{C}_\theta^\dagger= \boldsymbol\rho$), then the related single-particle reduced density matrix $\rho$ fulfills  Eq.(\ref{TH1}). 
If the quantum state $\boldsymbol\rho$ breaks $\mathcal{C}$-symmetry, i.e. if $\mathcal{C}^{}_\theta \boldsymbol\rho\mathcal{C}_\theta^\dagger \neq \boldsymbol\rho$ for any set of phases $\{ \theta \}$,   Eq.(\ref{TH1}) is also violated\cite{nota-gaussian-state},   and one can  introduce a quantifier for $\mathcal{C}$-symmetry breaking of the quantum state in terms of the single-particle density matrix. We shall consider two different quantifiers.
 
 {\it Global quantifier.} 
 The first natural option, straightforwardly suggested  by Eq.(\ref{TH1}), is to define
\begin{eqnarray}\label{QC-global-def}
Q^C_{glob} &=&  \frac{1}{\sqrt{2L}} \,\underset{\{ \theta\}}{\rm min} \,|| b^C (\{ \theta\}) || = \\
&=& \nonumber \frac{1}{\sqrt{2L}}  \,\underset{\{ \theta\}}{\rm min} \sqrt{\sum_{i,j=1}^L \sum_{\alpha,\beta=A,B} |b^C_{j \alpha,i \beta} (\{\theta\})|^2 } 
\end{eqnarray}
where
\begin{equation}\label{bC-def}
b^C (\{ \theta\})=    \rho+  \mathsf{U}_\theta^* \rho^* \, \mathsf{U}_\theta  -\mathbb{I} 
\end{equation}
is a Hermitean matrix in the single particle Hilbert space, $|| b^C||=\sqrt{(b^C)^\dagger b^C}$ is its norm, and the minimum is computed over all possible realizations of phases in Eq.(\ref{U-def}). 
Since $\mathsf{U}_\theta$ is diagonal in real space, the matrix $\mathsf{U}_\theta^* \, \rho^*   \,  \mathsf{U}_\theta$ appearing in Eq.(\ref{TH1}) differs from $\rho^*$ simply by additional phases  in the off-diagonal entries. 
Moreover, by noticing that $\rho+  \mathsf{U}^*_\theta \rho^* \, \mathsf{U}_\theta$ is a positively defined matrix, it can be shown that $Q^C_{glob}$ represents the minimal mean squared deviation of its eigenvalues  $\lambda^{(s)} (\{ \theta \})   \ge 0 $ (with $s=1,  \ldots 2L$) from the $\mathcal{C}$-symmetric case $\lambda^{(s)} (\{ \theta \}) \equiv 1$, i.e.
\begin{equation}
Q^C_{glob} =\underset{\{ \theta\}}{\rm min}\sqrt{\frac{1}{2L}\sum_{s=1}^{2L} \left(1-\lambda^{(s)}(\{ \theta \}) \right)^2}\quad.
\end{equation}
We shall dub such a quantifier {\it global}, for it involves the {\it entire} single-particle density matrix $\rho$, which can therefore by used to compute expectation values of any (local or non-local) one-body observables.

 {\it Local quantifier.} If, however, one is interested in measuring observables localized in one specific portion of the lattice, the knowledge of the entire $\rho$ is unnecessary, and only a sub-block of $\rho$ is needed. For definiteness, let us focus  e.g. on one lattice site   $P^*=(j^*,\alpha^*)$. In order to evaluate  the expectation values  of the local density $\langle {c}^\dagger_{P^*} {c}^{}_{P^*}\rangle$ at $P^*$,  of the inter-cell or intra-cell currents $\langle \hat{J}^{intra}_{j^*}\rangle $ and $\langle \hat{J}^{inter}_{j^*}\rangle $   across $P^*$ [see Eqs.(\ref{inter-cell-cur-def}) or (\ref{intra-cell-cur-def})], or also to compute the correlation  $\langle {c}^\dagger_{P^*} {c}^{}_{P}\rangle $ between $P^*$ and  any other lattice site $P=(j^* +l,\beta)$, only the $(j^*,\alpha^*)$-th row (or column) of $\rho$ is used. 
In particular, if the system is translationally invariant, the expectation value of a local observable is actually independent of the cell, and correlations only depend on the spatial cell distance, implying that the choice of the reference cell $j^*$ is irrelevant.
One can thus introduce a {\it local} $\mathcal{C}$-breaking quantifier as
\begin{eqnarray}\label{QC-local-def}
Q^C_{loc} &=& \,\underset{\{ \theta\}}{\rm min}| b^C_{j^*\alpha^*} (\{\theta\}) | =  \\
&=& \underset{\{ \theta\}}{\rm min} \sqrt{\sum_{i=1}^L \sum_{\beta=A,B} |b^C_{j^* \alpha^*,i \beta} (\{\theta\})|^2 } \quad.\nonumber
\end{eqnarray} 
A few remarks are in order. First, we note that both quantifiers (\ref{QC-global-def}) and (\ref{QC-local-def}) are defined upon minimization over all possible choices of the  $\{ \theta \}$ phases appearing in the definition of charge conjugation Eqs.(\ref{CC-def}) and (\ref{U-def}). Because each phase set $\{ \theta\}$ identifies a gauge transformation [see Eqs.(\ref{gauge-def}) and (\ref{0-to-g})], the minimizing phase set $\{ \theta \}$  also corresponds to the optimal gauge $[g]$, where the $\mathcal{C}_0$-symmetry is  broken  the least. This minimization is needed since $\mathcal{C}$-symmetry may be present, but hidden in a unsuitable choice of the gauge, like e.g. in the case where model (\ref{Ham-real}) is defined in OBCs.\\
Second, because expectation values of observables  must be gauge independent, they must be evaluated by writing both the state and the operator   in the same gauge, i.e. 
\begin{eqnarray}\label{gauge-indep}
    \langle \boldsymbol{O}\rangle= {\rm tr}[O \rho]={\rm tr}[{O}(\{\theta\})  {\rho}(\{\theta\})]
\end{eqnarray}
where  $\boldsymbol{O}=\sum_{I,J} O^{}_{IJ}{c}^\dagger_I c^{}_J=\sum_{I,J} ( {O} (\{\theta\})^{}_{IJ} \tilde {c}^\dagger_I \tilde{c}^{}_J$ is a one-body observable, and $I,J$ is a compact notation for the site,  $I=(i,\alpha)$ $J=(j,\beta)$.

\subsubsection{Chiral transformation} 
Proceeding similarly for the chiral transformation Eq.(\ref{S-def}), one can prove (see Appendix~\ref{AppA} for details) that, if  $\boldsymbol\rho$ is symmetric under the chiral transformation  ($\mathcal{S}  \boldsymbol\rho\mathcal{S}^\dagger= \boldsymbol\rho$), then the single-particle density matrix $\rho$ fulfills
\begin{equation}\label{TH-chiral}
 ( \rho- \mathbb{I} /2)+  \mathsf{U}_0  \,  ( \rho- \mathbb{I} /2) \,  \mathsf{U}_0  = 0 \quad,
\end{equation}
where $\mathsf{U}_0$ is given by Eq.(\ref{U0-def}).
Note that, differently from  Eq.(\ref{TH1}), the second term of Eq.(\ref{TH-chiral}) does not contain $\rho^*$. Introducing the matrix 
\begin{equation} \label{bS-def}
b^S  =    \rho+  \mathsf{U}_0  \rho \, \mathsf{U}_0  -\mathbb{I} \quad,
\end{equation}
the global and local quantifiers of chiral symmetry breaking are defined as
\begin{eqnarray}\label{QS-global-def}
Q^S_{glob}  =  \frac{1}{\sqrt{2L}}  \,|| b^S  || =  \frac{1}{\sqrt{2L}}   \sqrt{\sum_{i,j=1}^L \sum_{\alpha,\beta=A,B} |b^S_{j \alpha,i \beta}  |^2 } \,\,\,
\end{eqnarray}
and
\begin{eqnarray}\label{QS-local-def}
Q^S_{loc} &=& | b^S_{j^*\alpha}  | =   
    \sqrt{\sum_{i=1}^L \sum_{\beta=A,B} |b^S_{j^* \alpha,i \beta} |^2 } \quad.
\end{eqnarray} 
Notice that no minimization over the phase sets $\{\theta\}$ is needed for the quantifiers $Q_{glob}^S$ and $Q_{loc}^S$ because the chiral transformation  Eq.(\ref{S-def-pre}) is unaffected by gauge transformations (\ref{gauge-def}), due to its anti-linear character.

\subsection{Quantifiers for the ground state of the two band model Eq.(\ref{Ham-real})}
Let us now harness the  quantifiers introduced above to  evaluate the amount of  symmetry breaking on the ground state
 of the two-band model (\ref{Ham-real}).
In particular, we shall compare the effects of a local  $\mathcal{C}$-breaking  ($U \neq 0$ and $\varphi=0$) to the case of a non-local $\mathcal{C}$-breaking  ($U = 0$ and $\varphi \neq m \pi$). To this purpose we observe that, because the phase $\varphi$ appears divided by the number $2L$ of sites in the Hamiltonian (\ref{Ham-real}), for $L \gg 1$ each $\mathcal{C}$-breaking hopping term is of order $O(\varphi/L)$. In order to make a correct comparison, we thus require to have the same scaling law for the 
local $\mathcal{C}$-symmetry breaking induced by the on-site potential, and we shall  re-express $U$ in  Eq.(\ref{Ham-real}) as 
\begin{equation}
U=\frac{u}{L} \quad.
\end{equation}  
In this way, the non-dimerized case [$v=w$ in Eq.(\ref{Ham-real})] and the dimerized case [$v \neq w$] describe a   metal and an SSH insulator in the thermodynamic limit, respectively, with  a small  additional $\mathcal{C}$-breaking term  $O(\varphi/L)$ or $O(u/L)$. 
\begin{figure}[h]
\centering
\includegraphics[width= \columnwidth]{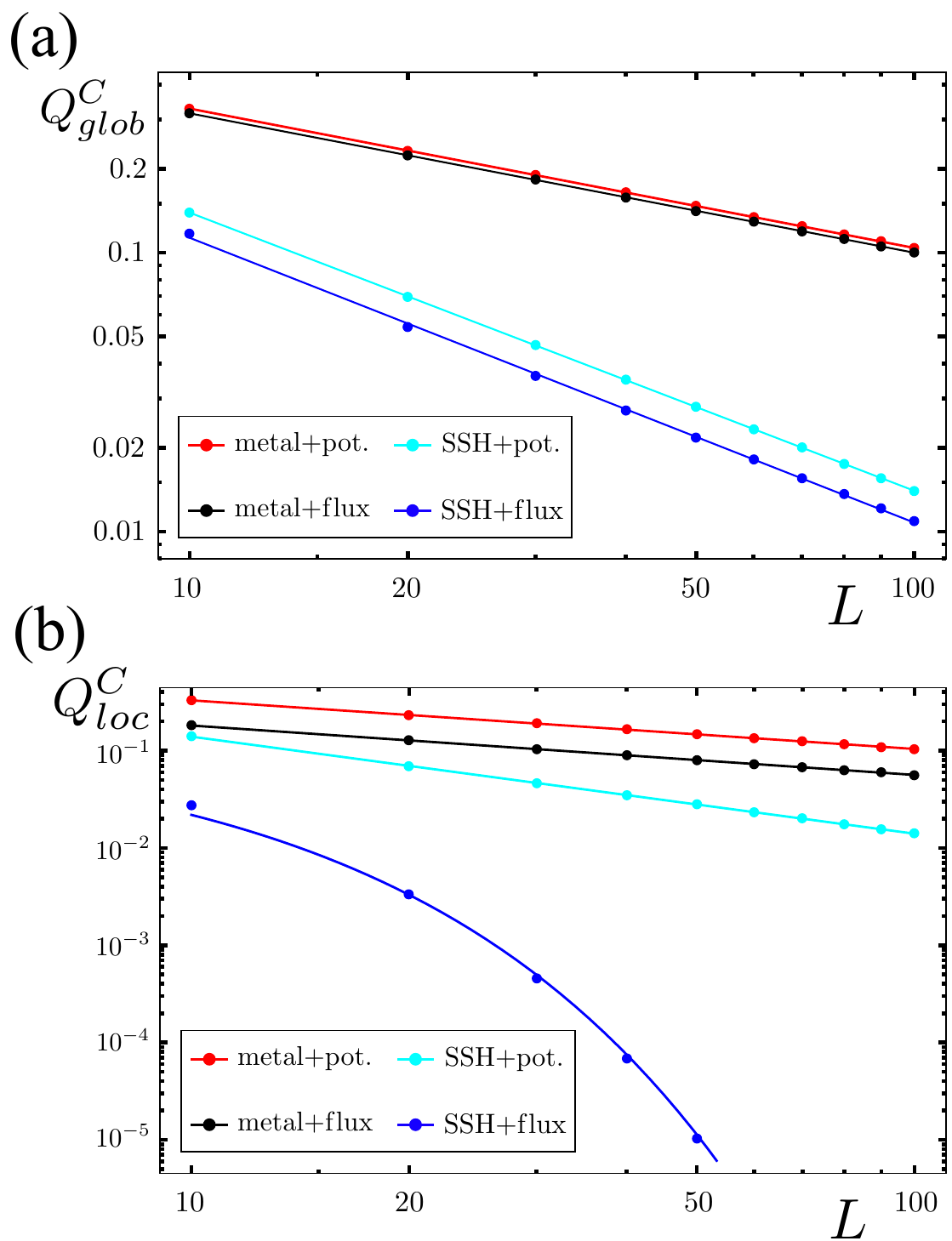}

\caption{\label{Fig5}  (a) The  global quantifier $Q^C_{glob}$ of $\mathcal{C}$-symmetry breaking, Eq.(\ref{QC-global-def}), is shown in log-log scale as a function of the number of cells $L$, for the  ground states of four different Hamiltonians: The  metallic case ($v=w$) and the insulating SSH case ($v=1$ and $w=0.7$), where the $\mathcal{C}$-symmetry is broken locally by an on-site potential $u=1$ (red curve and cyan curves, respectively), and broken non-locally by a flux $\varphi=\pi/2$ (black and blue curves, respectively).
(b) The behavior of the local quantifier $Q^C_{loc}$ of $\mathcal{C}$-symmetry breaking, Eq.(\ref{QC-local-def}), is shown in log-log scale as a function of $L$, for the same ground states as in panel (a). For the case of the SSH insulator with $\mathcal{C}$-symmetry broken non-locally by the flux (blue curve), $Q^C_{loc}\sim\exp[-L/\Lambda]$ decreases exponentially with the system size $L$ (here $\Lambda\simeq 5.3$), in striking contrast with the algebraic decay of the other local quantifiers and of all  global quantifiers.
   }
\end{figure}

We have first considered the global $\mathcal{C}$-symmetry breaking quantifier $Q^C_{glob}$ in Eq.(\ref{QC-global-def}). The minimization over the phase set $\{ \theta\}$,  performed over various physically reasonable  phase choices\cite{nota-minimization}, leads to the result shown  in  Fig.\ref{Fig5}(a), where the global quantifier $Q^C_{glob}$ is plotted for all these cases as a function of the systems size $L$. All curves in the log-log plot exhibit a  linear behavior, indicating that    $Q^C_{glob}$ is always  decaying {\it algebraically} with the system size $L$, with a power law determined by the slope. In particular, $Q^C_{glob}\sim 1/\sqrt{L}$ in the metallic case (red and black curves), while  $Q^C_{glob} \sim 1/L$ is found for  the insulating SSH case (cyan and blue curves), regardless of whether the symmetry is broken locally by the staggered potential $U$ or non-locally by the flux $\varphi$.

Let us now turn  to the local quantifier in  Eq.(\ref{QC-local-def}). For definiteness, we have chosen $P^*=(L/2,A)$  as a reference site, i.e. the   site  $\alpha^*=A$ in the ``central" cell $j^*=L/2$, and we have computed how $Q^C_{loc}$ scales with $L$ for the various cases,   as displayed in Fig.\ref{Fig5}(b). For a metallic ground state where the $\mathcal{C}$-symmetry  is broken locally (red curve) and non-locally (black curve), $Q^C_{loc}$ still exhibits an algebraic decay $Q^C_{loc}\sim 1/\sqrt{L}$ with the system size. Similarly, for the   insulating SSH ground state an algebraic decay $Q_{loc}^C \sim 1/L$ is again observed when the $\mathcal{C}$-symmetry is broken locally by the on-site staggered potential (cyan curve). Thus, in all these 3 cases  the local quantifier $Q_{loc}^C$  in  Fig.\ref{Fig5}(b) exhibits the same scaling behavior  as the corresponding global quantifier $Q_{glob}^C$ shown in Fig.\ref{Fig5}(a).
However, in the case of an  insulating SSH ground state where  the $\mathcal{C}$-symmetry  is broken non-locally by the flux,  one finds an {\it exponential} decay for the local quantifier $Q^C_{loc} \sim \exp[-L/\Lambda]$ with $\Lambda$ depending on the dimerization parameter $r$, as shown by the blue curve in  Fig.\ref{Fig5}(b).

The strikingly different
scaling law of $Q^C_{loc}$ obtained in the case of SSH model with flux, as compared to both $Q^C_{loc}$ in the other cases  and to the global   quantifier  $Q^C_{glob}$ in all cases,
provides a quantitative characterization of the insulator local rigidity to the flux from the perspective of charge conjugation symmetry breaking.\\

At a mathematical level,   this difference originates from the different phase sets $\{ \theta\}$ that minimize  the quantifiers  Eqs.(\ref{QC-global-def}) and (\ref{QC-local-def}) in the various cases. While these technical aspects are discussed in details   in Appendix~\ref{AppD}, here we point that the local insensitivity of insulators to a non-local symmetry breaking induced by the flux  can be understood in more physical terms as follows. Because expectation values $\langle \boldsymbol{O} \rangle$ are independent of the gauge choice, 
when considering  observables $\boldsymbol{O}_{P^*}$ localized around (say) an arbitrarily chosen site $P^*=(j^*,\alpha^*)$, it is always possible to evaluate $\langle \boldsymbol{O}_{P^*}\rangle$  in  the gauge sketched in Fig.\ref{Fig3}(b),  where the flux is accumulated along a link located farther than the localization length $\lambda$ from $P^*$. This is realized by inserting  in Eq.(\ref{gauge-def}) the phase set $\{ \theta_\ell \}$ given in Eq.(\ref{theta-ell}). In such a gauge, the  $(j^*,\alpha^*)$-th row of the single-particle density matrix ${\rho}(\{\theta_\ell\})$, which contains the actual entries involved in the evaluation of $\langle \boldsymbol{O}_{P^*}\rangle$ through Eq.(\ref{gauge-indep}), becomes effectively independent of the flux phase $\varphi$, due to the finite localization length of the insulator [see Fig.\ref{Fig8}(b) in App.\ref{AppD}].
 At the same time,  any  possible spur of $\varphi$  also  disappears  from the expression ${O}_{P^*}(\{\theta_\ell\})$ of the local operator in that gauge, precisely because it is localized around $P^*$ only. This is  the case for  the on-site density operator, but also e.g. for the current operator across $P^*$,  where the flux phase appearing in the native gauge [see Eqs.(\ref{inter-cell-cur-def}) in (\ref{inter-cell-cur-def})]   is removed  by the gauge transformation from the neighborhood of $P^*$. In conclusion $\langle \boldsymbol{O}_{P^*}\rangle$ is independent of the flux.

\subsection{Dynamical evolution of the quantifiers}
There is one further interesting advantage provided by the introduced symmetry breaking quantifiers $Q_{glob}$ and $Q_{loc}$. Because they are   functions of the   single-particle density matrix $\rho$ of the system, as   $\rho$ dynamically evolves, one can monitor how the amount of symmetry breaking changes with time when the system is driven out of equilibrium. 
To this purpose, some care must be taken for the charge conjugation quantifiers, though. 
Indeed, if the minimization over the phase set $\{\theta \}$ in Eqs.(\ref{QC-local-def}) and (\ref{QS-local-def}) were  performed at every time, the minimizing set $\{\theta \}(t)$ would     change the definition of charge conjugation transformation $\mathcal{C}_{\theta(t)}$ on the run [see Eqs.(\ref{CC-def})-(\ref{U-def})]. This would have   serious implications, for it would 
lead to a time-dependent gauge transformation   [see Eq.(\ref{0-to-g})] and thereby introduce  a vector potential in the Hamiltonian (\ref{Ham-real})  describing a spurious space and time-dependent electric field, with vanishing circulation along the ring.
In order to avoid such an unphysical effect, one has thus to first determine  the phase set $\{\theta \}^i$ minimizing   $Q^C_{glob}$ and $Q^C_{loc}$ for the density matrix of the initial state. This identifies the definition of  charge conjugation transformation  $\mathcal{C}_{\theta^i}$ that is the least broken at $t=0$. Then, by  keeping such phase set frozen, i.e. by consistently retaining the definition of $\mathcal{C}_{\theta^i}$, one can inserting the dynamically evolving $\rho(t)$ into $b^C(\{ \theta^i\})$ in Eq.(\ref{bC-def}), and thus evaluate  how  $|b^C_{j^*\alpha^*}(\{ \theta^i\})|$ changes with time.
This is how we determined the red curve in Fig.\ref{Fig4}(d), which shows that the $\mathcal{C}$-symmetry   remains effectively unbroken until the time $\tau^*_1$, causing the tsunami effect.
In contrast, the chiral symmetry breaking quantifier $Q_{loc}^S$ does not require any minimization over $\{ \theta \}$ by definition, as observed in Sec.\ref{sec5a}. The dynamical evolution is thus straightforwardly obtained by inserting $\rho(t)$ into Eqs.(\ref{bS-def}) and (\ref{QS-local-def}). The obtained green curve   in Fig.\ref{Fig4}(d)  clearly shows  that in a quenched SSH insulator threaded by a flux  the chiral symmetry is dynamically broken immediately after the quench, whereas this is not the case for the charge conjugation, which remains effectively unbroken up to the extensively long time $\tau_1^*$, i.e.  $t=L \hbar /4 \gamma \mathsf{v}_m$.

We  conclude by observing   that, because   the  symmetry breaking quantifiers are  based on the single particle reduced density matrix $\rho$ of the quantum state, they  are closely related to measurable quantities. Indeed the diagonal entries of $\rho$ describe the local particle density and can be directly measured, while its off-diagonal entries encode all two-point correlation functions, which can  be experimentally   detected   with various techniques, 
 e.g.   quantum gas microscopy\cite{bloch-kollath,gross-bakr} and matter wave interferometry\cite{schmiedmayer}.
 The symmetry breaking quantifier sets an upper bound for  the deviations of  density or correlation functions from the symmetric case. This is shown precisely in Fig.\ref{Fig4}: As long as the $\mathcal{C}$-breaking quantifier  remains negligible    [red curve in panel (d)], the correlations and the density deviation do not feel the effect of the $\mathcal{C}$-breaking flux [see panels (a) and (b)]. Only after the quantifier has become  a significant fraction of unity, can these quantities  start to deviate from the zero-flux case.\\\\

\section{Robustness of the effect}
\label{sec6}
While  in Sec.\ref{sec4} the tsunami effect has been explicitly demonstrated  in a clean and non-interacting system, we shall now relax these hypotheses and address the generality and robustness of this effect by studying its stability against disorder  and   interactions.
 
\subsection{Effects of disorder}
Let us start by introducing some disorder in the tunneling amplitudes of Eq.(\ref{Ham-real}). We shall thus consider the disordered SSH Hamiltonian 
\begin{eqnarray}
  \lefteqn{   {H}^{dis}_{SSH}  = }  & &\label{Ham-real-dis}\\
  &=&\gamma  \sum_{j=1}^{L}   \left( v_j \,e^{i \frac{\varphi}{2L} }{c}_{j A}^\dagger {c}^{}_{j B} + w_j \, e^{i\frac{\varphi}{2L} }{c}_{j B}^\dagger {c}^{}_{j+1 A} + \text{H.c.} \right) \nonumber
    \end{eqnarray}
with
\begin{equation} \label{disorder}
\begin{array}{lcl}
v_j &=& v (1+ \sigma\, \xi_j) \\
w_j &=& w (1+ \sigma\, \eta_j)
\end{array}
\end{equation} 
Here $v$ and $w$ represent the average   magnitude of the tunneling amplitudes, $\sigma$ is a parameter identifying the disorder strength, and $(\xi_j, \eta_j) \in [-1/2; 1/2]$, with $j=1\ldots L$, are $2L$ uniformly distributed random numbers. Notice that disorder in the tunneling does not alter the $\mathcal{C}$-symmetry properties with respect to the clean case, and the Hamiltonian (\ref{Ham-real-dis}) exhibits charge conjugation symmetry for $\varphi=\pi m$ with $m\in \mathbb{Z}$.  
\begin{figure}[h]
\centering
\includegraphics[width=  \columnwidth]{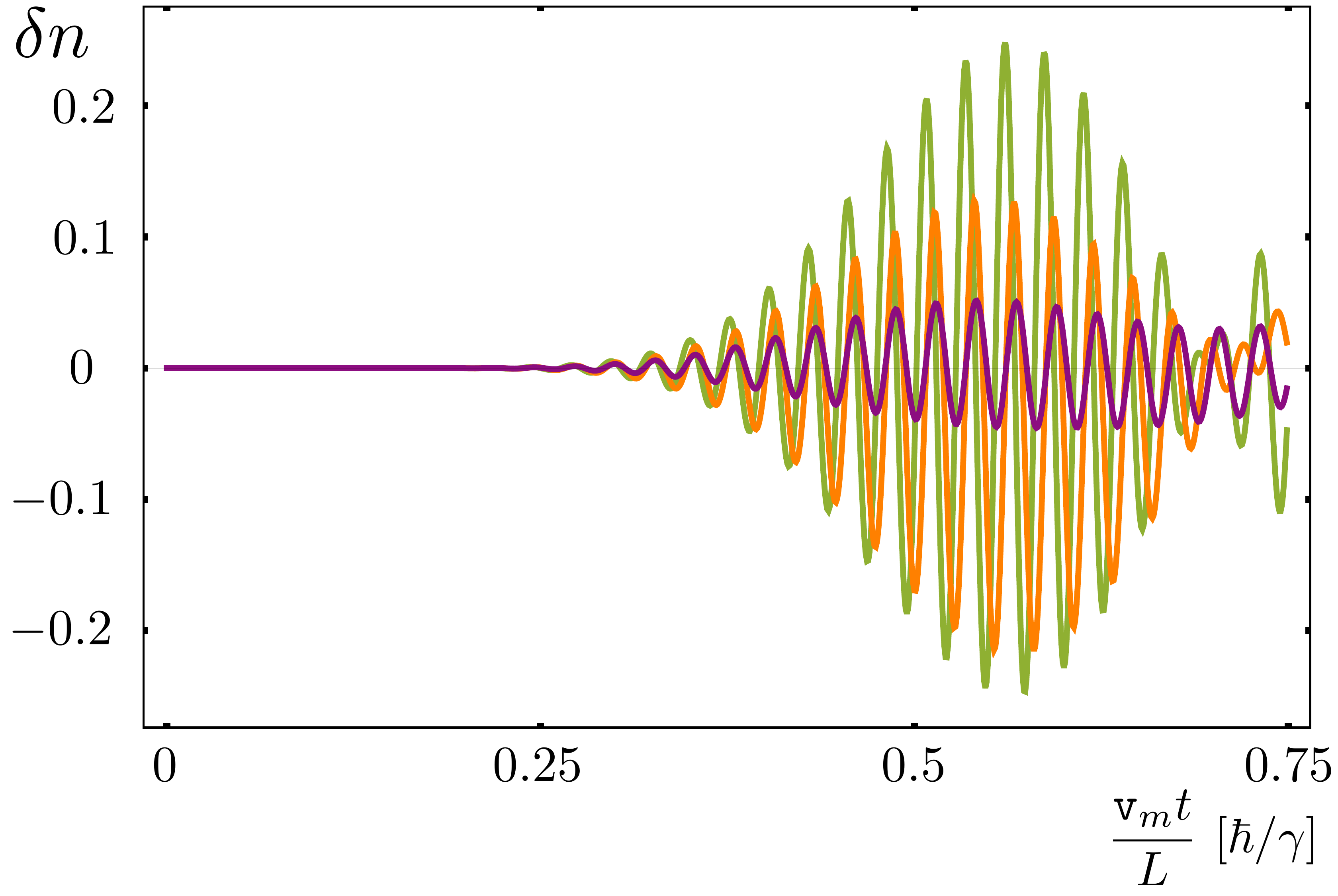}
\caption{\label{Fig6}  Effects of disorder (\ref{disorder}). The dynamical behavior of $\delta n$, the local density deviation from $1/2$, is plotted as a function of time for a quenched SSH ring,   for various values of the disorder  strength $\sigma$ [see Eq.(\ref{disorder})], namely  $\sigma=0$ (clean case, green curve),  $\sigma=0.1$ (orange curve) and   $\sigma=0.2$ (purple curve).  The number of cells is $L=12$  ($=24$ sites), the dimerization is $r=0.1$, the flux phase is $\varphi=\pi/2$ and  quench protocol is specified in Eq.(\ref{quench}). }
\end{figure}

We perform a quench in the average values $v$ and $w$ of Eq.(\ref{disorder})   according to the protocol (\ref{quench}) and analyze the dynamics of the local density. Due to the lack of translational invariance, one  has $n_{j, A/B}=1/2 \pm \delta n_{j}$, where  the local deviation $\delta n_{j}$ from half filling now depends on the cell, in general.  
In Fig.\ref{Fig6} we report the dynamical behavior of $\delta n$ on a given $A$ site, which is qualitatively well representative of the result for any site, and also of the spatial average over all   $A$ sites (not shown here). While the green curve refers to the clean case ($\sigma=0$) and is plotted as a reference, the orange and purple curves describe the effects of disorder  strength $\sigma=0.1$, and $\sigma=0.2$, respectively. As one can see, while the disorder strength   reduces the amplitude of the tsunami effect, the qualitative behavior and the extensively long time needed for its appearance remain unaffected.

In order to understand this effect, we recall that any finite amount of disorder in the tunneling amplitudes of a 1D system is sufficient to localize the single particle eigenstates of the Hamiltonian in the thermodynamic limit\cite{anderson_1958,lee_1985}.
However, in the case of a finite system, the  localized {\it vs} extended  nature of the single particle eigenstates effectively depends on the ratio between localization length and system size.
In particular, since the localization length depends on the single particle energy, a fraction of the single particle eigenfunctions might   extend  over the entire system for sufficiently weak disorder.
Hence, the system behaves as if a fraction of the particles could still propagate throughout the ring and reveal the existence of the flux once the extensive time has elapsed.

\subsection{Effects of interaction}
Let us now probe the stability of the tsunami effect against particle-particle  interaction.  In particular, we shall add to the SSH Hamiltonian [Eq.(\ref{Ham-real}) with $U=0$] the following term
\begin{eqnarray}\label{Hint}    
     {H}^{int}  = \gamma \, V  \sum_{j=1}^L   ( {n}_{j A}+{n}^{}_{j+1 A}-1)  ( {n}^{}_{j B}-\frac{1}{2} )  
\end{eqnarray}
describing a nearest neighbor interaction between particle density fluctuations $(n_{j\alpha}-1/2)$ from the half filling value $1/2$. Here $V$ is the dimensionless coupling constant expressing the interaction strength in units of the energy unit $\gamma$ appearing
in Eq.(\ref{Ham-real}).
Importantly,  the interaction term fulfills $\mathcal{S}^{}   {H}^{int}  \mathcal{S}^\dagger ={H}^{int}$ (chiral symmetry) and $\mathcal{C}_\theta^{}   {H}^{int}  \mathcal{C}^\dagger_\theta={H}^{int}$ (charge conjugation symmetry) for any choice of $\theta$-phases in Eqs.(\ref{CC-def})-(\ref{U-def}), so that  the total Hamiltonian \begin{eqnarray}
{H}  &=& \gamma  \sum_{j=1}^{L}   \left( v \,e^{i \frac{\varphi}{2L} }{c}_{j A}^\dagger {c}^{}_{j B} + w \, e^{i\frac{\varphi}{2L} }{c}_{j B}^\dagger {c}^{}_{j+1 A} + \text{H.c.} \right) + \nonumber \\
& & + {H}^{int} \label{Htot}
    \end{eqnarray}
still fulfills the chiral symmetry, while  the charge conjugation symmetry   is only broken by the global constraint imposed by a finite flux in the non-interacting SSH term.

\begin{figure}[h]
\centering
\includegraphics[width=  \columnwidth]{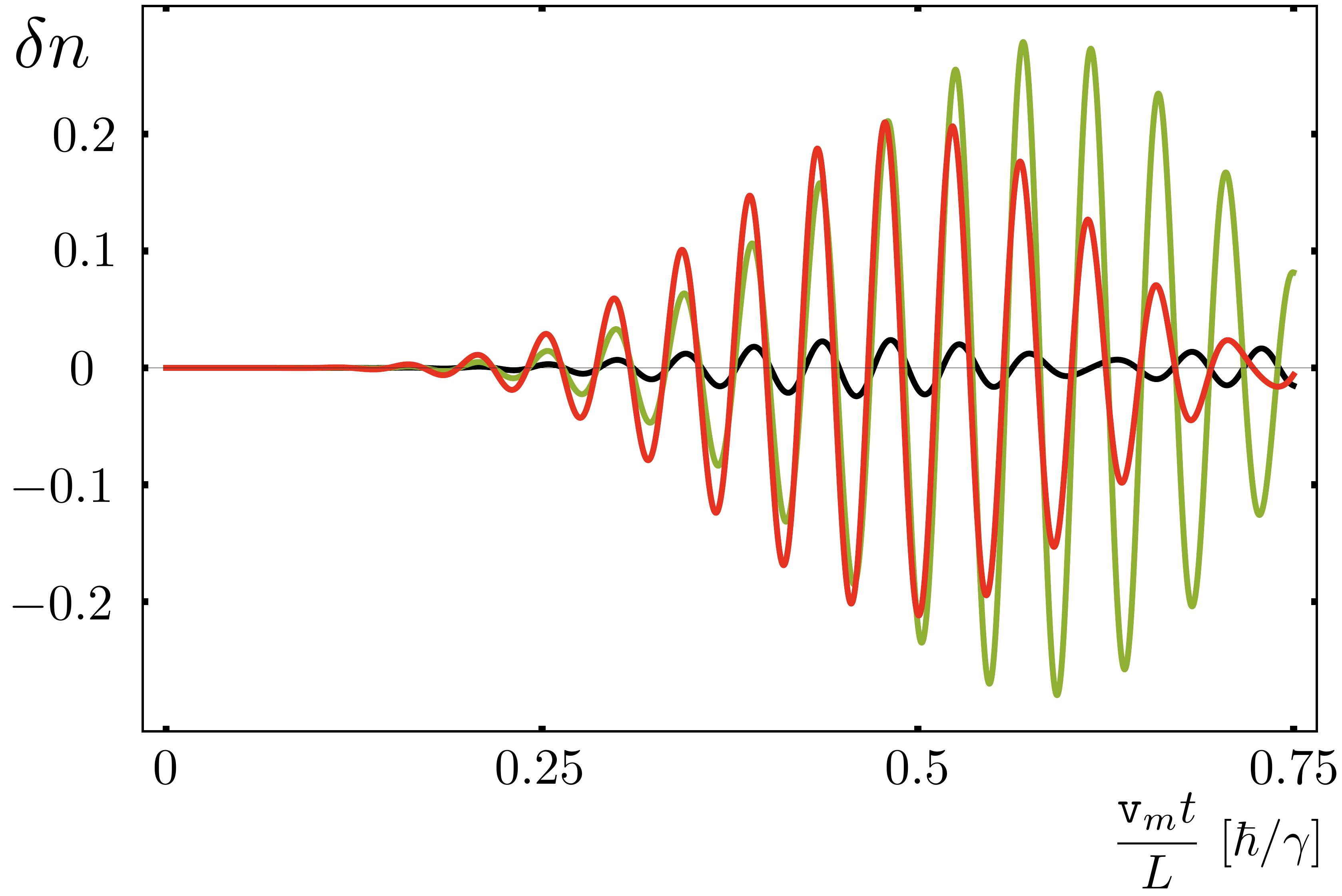}
\caption{\label{Fig7} Effects of interaction term Eq.(\ref{Hint}). The dynamical behavior of the local density deviation $\delta n$ is plotted as a function of time for a SSH ring with $L=7$ cells ($=14$ sites), threaded by a flux ($\varphi=\pi/2$) and undergoing a quench (\ref{quench}) with dimerization $r=0.1$, for various values of the   interaction strength $V$, namely  $V=0$, (green curve),  $V=+0.1$ (red curve) and   $V=-0.1$ (black curve).
 }
\end{figure}

For sufficiently strong interaction strength $V$, exceeding a critical value $V_c$, the half filled ground state of the   model (\ref{Htot}) is known to exhibit  a   transition  from a   band insulator to an interaction-induced insulating phase. In particular, a   repulsive interaction  $V>V_c>0$  leads to degenerate charge-density wave  phases characterized by a uniform density imbalance between $A$ and $B$ sublattices. By contrast,
 an attractive interaction $V<-V_c<0$ leads to a regime of  degenerate states characterized by phase separation,  where fermions equally populate the A and B sublattices while remaining compressed in one half of the system\cite{Fraxanet2022}.   The critical value $V_c$ of the transition depends on the dimerization strength and on the sign of interaction, e.g. for $r=1/3$ one has $V_c=4$ for the repulsive case, while $V_c\simeq  0.87$ for the attractive case\cite{Fraxanet2022}.

While the analysis of the dynamical effects of strong interaction is far beyond the purposes of the present work, one can argue that, because the ground state on the non-interacting model is insulating, the inclusion of  sufficiently weak interactions ($|V| \ll V_c$), leads to an interacting
   ground state that is  non-degenerate and adiabatically connected to the non-interacting case.   Moreover, when the quench is applied, 
the velocity of the resulting excitations is expected to be sightly enhanced by a repulsive interaction and slightly reduced by an attractive one \cite{voit-review,barbiero2017}.  
  Because such velocity  determines the spreading of correlations and thereby the onset of the tsunami effect, we expect interactions to  modify quantitatively, but not qualitatively, the  phenomenon. 
 In order to test these expectations, we have adopted a
 numerically exact diagonalization method  to determine the initial interacting ground state and to investigate the out-of-equilibrium behavior governed by the Hamiltonian~(\ref{Htot}), under the quench protocol (\ref{quench}). 
 While being limited to small system sizes (here $L=7$, i.e. $14$ sites), this approach allows us to explore the dynamical evolution  at arbitrary long times.  
 
The  result is shown in Fig.\ref{Fig7}, which displays the dynamical evolution of the density deviation $\delta n$ for a quenched   interacting SSH model with dimerization $r=0.1$, threaded by a flux $\varphi=\pi/2$, and for three values of interaction $V$. The green curve refers to the non-interacting case ($V=0$) and is meant as a reference case, while the red and black curves describe the cases of weakly repulsive ($V=0.1$) and weakly attractive interaction ($V=-0.1$), respectively.
As one can see,    the density deviation $\delta n$ still remains extremely small for an extensively long time, after which it starts to take seizable values, therefore indicating that the  tsunami effect is not destroyed by weak interactions. Also,  an inspection of the density oscillation maxima shows that in the repulsive (attractive) case the onset of the tsunami effect occurs slightly earlier (later) than the non interacting case, in agreement with our expectations. We note, however, that attractive interactions tend to suppress the magnitude of the fluctuations more than the repulsive case. This seems to be the dynamical counterpart of what happens for the ground state, where, for   given dimerization $r$ and interaction magnitude $|V|$ of interaction, the attractive case is closer to the   critical value $V_c$ than the repulsive case.

\section{Discussion and  conclusions}
\label{sec7}
We have investigated the quantum dynamics of a quenched fermionic system on a 1D dimerized ring-shaped lattice with PBCs.
By analyzing the effects of an explicit breaking of the charge-conjugation symmetry $\mathcal{C}$ on the local observables and correlations, we have shown that   the  impact of the  symmetry breaking in the Hamiltonian is felt by local observables in a   way that heavily depends on the localization properties of the initial quantum state and on the local {\it vs} non-local nature of the $\mathcal{C}$-breaking.
In particular, if the $\mathcal{C}$-symmetry is broken non-locally by a magnetic flux and the localization length of the quantum state is finite (insulator), local observables behave is if $\mathcal{C}$-symmetry were unbroken, in striking contrast with what happens in metals or in insulators with $\mathcal{C}$-symmetry  broken locally by an on-site potential.

While at equilibrium or in adiabatically slow dynamics, the rigidity of an insulator to a flux was known, the most spectacular effect that we found is that such rigidity to the $\mathcal{C}$-symmetry breaking persists even in far from equilibrium conditions, i.e. when a quantum quench is applied.  In particular, local observables effectively retain their $\mathcal{C}$-symmetric values  
  for a time that scales linearly with  the system size, and only after such time can  the effects of non-local $\mathcal{C}$-symmetry breaking become visible. 
A clear evidence of this phenomenon, that we have dubbed  tsunami effect, is shown in Fig.\ref{Fig1}, which displays the dynamical evolution of the local particle density deviation $\delta n$ from $1/2$ resulting from a quench, and reveals the striking difference between   local $\mathcal{C}$-symmetry breaking (RM model with on-site potential) and non-local $\mathcal{C}$-symmetry breaking (SSH model with flux). In the strongly dimerized limit, we have also been able to determine an  analytical expression  for the dynamical behavior of $\delta n$ [see Eq.(\ref{deltanA-asympt}) and Fig.\ref{Fig2}], characterized by an exponential suppression as a function of the system size~$L$  before the tsunami onset  [see Eq.(\ref{expo})]. As discussed in Sec.\ref{sec4B}, the tsunami effect is not limited to the quench protocol (\ref{quench}) and to the strongly dimerized limit. It is a general phenomenon occurring  for any quench protocol, provided that the $\mathcal{C}$-symmetry is broken by a flux and the initial state is insulating.  In contrast, when the initial state is metallic, the deviation  $\delta n$ appears as $O(1/L)$ right after the quench [see Eq.(\ref{deltan-metal})].

The long time invisibility of the $\mathcal{C}$-breaking flux at local level is a physical effect that is of course completely independent of the gauge chosen for the vector potential. However, in each arbitrary point where e.g the density is evaluated, it can  most suitably understood in the gauge sketched in Fig.\ref{Fig3}(b), where the entire flux phase is accumulated
to opposite side of the ring.

Moreover, the tsunami effect also manifests itself in the correlation function, as displayed in Fig.\ref{Fig4}(a). In fact,    the earliest manifestation of the   flux occurs in the non-local correlations of two maximally separated points ($l=\pm L/2$), again only after an extensive time $\tau^*_1$ in Eq.(\ref{tau-1-star}). 
As shown in Fig.\ref{Fig4}(c), this time can also be seen as the moment when the Wannier function localization length $\lambda$, which is spreading as a result of  the quench, reaches its maximal value  and has thus explored to entire ring, thereby experiencing the presence of the non-local flux. 
As highlighted by red and black vertical dashed lines in Fig.\ref{Fig4}, the time $\tau_2^*$ where the tsunami effect is perceived on the density, i.e. on local correlations at $l=0$, is twice longer than $\tau_1^*$.\\

In order to characterize this phenomenon, we have also introduced the quantities $Q_{glob}$ and $Q_{loc}$, which quantify how strongly a symmetry is broken in an arbitrary quantum state by means of its single particle density matrix, at global and at local level, respectively. This has enabled us to obtain two results. First, as far as the equilibrium ground state is concerned,   the known insensitivity of insulator to a flux can be  interpreted as a local rigidity to the $\mathcal{C}$-symmetry breaking.
Explicitly,    the ``amount" $Q_{loc}^C$ of local $\mathcal{C}$-symmetry breaking of a SSH ring threaded by a flux is found to be exponentially small in the system size $L$, while it only exhibits an algebraic decay in insulators with local $\mathcal{C}$-symmetry breaking and in metals (see Fig.\ref{Fig5}). We have also shown that this can be straightforwardly understood in a suitable    gauge, as illustrated in Fig.\ref{Fig3}(b).
 Second, in terms of out of equilibrium dynamics, the  obtained evolution of the local  quantifiers $Q^C_{loc}$ and $Q^S_{loc}$   has allowed us to demonstrate 
 that the tsunami effect, i.e. the predicted robustness to charge conjugation symmetry breaking,  is completely unrelated to the presence of the chiral symmetry, which is instead lost right after the quench because of its dynamical breaking, as shown in Fig.\ref{Fig4}(d).  
 Furthermore, by including  weak disorder and density-density interaction, we have proven that  the tsunami effect is   not destroyed by these effects (see Fig.\ref{Fig6} and Fig.\ref{Fig7}). \\

{\it Space-time scaling regime.} We also emphasize that   
our work represents   further advances in the exploration of the space-time scaling regime, where the out of equilibrium quantum dynamics is analyzed for timescales $t$ that are proportional to the system size $L$.
It was recently   shown that this regime offers a different perspective to identify out of equilibrium topological classification of quantum systems\cite{rossi-budich-dolcini}, as compared to  the more conventional short time limit\cite{spielman_RMP_2019} and adiabatic limit.\cite{rigol_2017,pollman_2017,vishwanath_2017,keesling_2019,barbarino_2020}
 In terms of the parameter $\eta=2\pi t /L$ that identifies the ratio between time and system size (see Ref.[\onlinecite{rossi-budich-dolcini}]), the standard thermodynamic limit corresponds to $\eta \rightarrow 0$. In this limit, the system cannot experience the presence of the flux and it behaves as if charge conjugation were effectively unbroken. In contrast, at finite values of $\eta$, non trivial effects emerge. Specifically, by re-expressing our result Eq.(\ref{deltanA-asympt})  in terms of $\eta$, the tsunami effect can be regarded to as a {\it dynamical crossover} characterizing the density deviation  $\delta n$. Indeed, for $\eta < \eta_c $, where $\eta_c=\pi \hbar/\gamma \mathsf{v}_m$, we find an exponential suppression $\delta n \sim e^{-\kappa(\eta) L}$ as a function of the system size, typical of an insulator [with   $\kappa(\eta\rightarrow \eta_c)=0$]. In contrast, for $\eta>\eta_c$, one finds the slow algebraic decay $\delta n \sim 1/\sqrt{L \eta^{3}}$ with $L$, typical of a metal,
so that the quantum non locality induced by a flux appears. Our result thus proves that only in the space-time scaling regime one can  observe  such ``insulator-to-metal" dynamical crossover.

{\it Experimental realizations.} Finally, we would like to discuss some  possible strategies to implement the model in Eq.(\ref{Ham-real}) and to probe its out-of-equilibrium behavior. 
Although dimerized lattice models have been realized  with cold atoms in optical lattices in various setups  over the last decade\cite{bloch_2013,gadway_2016,gadway_2019,spielmann_2022,esslinger_2022}, most of these implementations are based on lattices with OBCs, whereas a crucial aspect for the observation of the tsunami effect is  a ring-shaped geometry with PBCs  and a flux. Recent experimental advances offer  promising perspective to  realize this type of setup. A first  proposal is based on Rydberg atoms in optical tweezers\cite{browaeys2020}, which allow one to engineer spin models equivalent to the fermionic Hamiltonians discussed here, also with dimerization\cite{leseleuc_2019}. In particular, in recent experiments   based on twezeer engineering, a tunable flux  was obtained by exploiting the synthetic dimension of Rydberg atoms \cite{lienhard2020,chen2023}, and ring-shaped geometries have also been recently designed\cite{piercivalle2022}. A further experimental platform suitable to test our findings could possibly be based on ${\rm SU}(N)$ ultracold fermions in optical lattice \cite{pagano2014,zhang2014,scazza2014}. In that case, effective rings can be created in synthetic dimension by coupling the different internal states of the specific atomic species. More precisely, the use of Raman lasers enables one to convert an internal state into a different one, thereby mimicking effective tunneling processes, with a twofold advantage. On one hand the couplings between different internal states are tunable, thus making possible to achieve effective dimerizations. On the other hand, complex tunneling processes can be easily engineered\cite{dalibard-review_2011,spielman_RPP_2014}, in order to mimic the effect of a magnetic flux. Finally, quantum gas microscopy\cite{gross-bakr}, widely used in tweezer and optical lattice setups, allows for an accurate detection of the local density and therefore for an accurate probing of the   effects predicted here.

\acknowledgments
F.D. acknowledges  financial support from the Italian Centro Nazionale di Ricerca in High Performance Computing, Big Data and Quantum Computing, funded by European Union - NextGenerationEU (Grant No. CN00000013).  J.C.B. acknowledges financial support from the German Research Foundation (DFG) through the Collaborative Research Centre SFB 1143 (Project No. 247310070), the Cluster of Excellence ct.qmat (Project No. 390858490), and the DFG Project 419241108

\appendix
\section{Proof of the Theorem}
\label{AppA}
In this Appendix we prove the theorem stated in Sec.\ref{sec2a}. Let us  assume that the many-particle state $\boldsymbol{\rho}$ of the system is invariant under charge conjugation, i.e. $\mathcal{C} \boldsymbol{\rho} \, \mathcal{C}^\dagger = \boldsymbol{\rho}$.
In order to prove  Eq.(\ref{TH1}),  we introduce the multi-index $J=(j,\alpha)$ to label the cell $j$ and the site $\alpha=A,B$, 
and observe that the charge conjugation transformation (\ref{CC-def}) can be written as
\begin{equation}
\mathcal{C} {c}^\dagger_{I}\mathcal{C}^\dagger= \sum_N \mathsf{U}_{ I N} {c}_{N}
\end{equation}
where  $\mathsf{U}$ is the matrix  written in Eq.(\ref{U-def}). In the following proof, we shall actually be slightly more general, and derive the relation of the single particle density matrix for an arbitrary unitary matrix $\mathsf{U}$, whence Eq.(\ref{TH1}) will follow as a particular case. We thus 
 observe that  the expectation value of any one-body observable $\boldsymbol{O}=\sum_{I,J} O^{}_{IJ}{c}^\dagger_I c^{}_J$ can be written as  
\begin{eqnarray}
\langle \boldsymbol{O}\rangle  =    \text{Tr} \{ \boldsymbol{\rho} \, \boldsymbol{O} \} =  \sum_{I,J}      O_{IJ} \text{Tr} \{ \boldsymbol{\rho} \,{c}^\dagger_I {c}^{}_J \} =    \text{tr}\{ O  \rho \} \label{equality-1}
\end{eqnarray}
where  $\rho$  is the single particle reduced density matrix  and the symbols ``${\rm Tr}$" and ``${\rm tr}$" denote traces over the Fock space and the single-particle Hilbert space, respectively.
At the same time, because of the $\mathcal{C}$-symmetry, the expectation value can also be written as (sum over the repeated indices is implicit here below)
\begin{eqnarray}
\lefteqn{\langle \boldsymbol{O}\rangle  = \text{Tr} \{ \mathcal{C}^\dagger \boldsymbol{\rho} \, \mathcal{C} \boldsymbol{O} \} =    O_{IJ} \text{Tr} \{  \boldsymbol{\rho} \, \mathcal{C} {c}^\dagger_I \mathcal{C}^\dagger \mathcal{C} {c}^{}_J \mathcal{C}^\dagger  \} = } && \nonumber \\
  &=&   O_{IJ} \mathsf{U}_{IN} \mathsf{U}_{JM}^* \text{Tr} \{  \boldsymbol{\rho} \, {c}^{}_N  {c}_M^\dagger  \}=  \nonumber \\
     &=&   O_{IJ} \mathsf{U}_{I N} \mathsf{U}_{JM}^* \text{Tr} \{  \boldsymbol{\rho}( \, \delta_{NM} -  {c}_M^\dagger {c}^{}_N  )\} =  \nonumber\\
     &=&   O_{IJ} \mathsf{U}_{I N} \mathsf{U}_{JM}^* (  \delta_{NM} -  \rho_{NM}  ) =  \nonumber \\
     &=&    O_{IJ}  \left( \delta_{IJ}   -  \left[ \mathsf{U} \rho \, \mathsf{U}^\dagger \right]_{IJ} \right) =     O_{IJ}  \left( \delta_{IJ}   -  \left[ \mathsf{U} \rho \, \mathsf{U}^\dagger \right]^*_{JI} \right) =  \nonumber \\
  &=& \text{tr}\left\{  O  \left[\mathbb{I} - \left( \mathsf{U} \rho \, \mathsf{U}^\dagger \right)^* \right] \right\} \quad. \label{equality-2}
\end{eqnarray}
Because both equalities (\ref{equality-1}) and (\ref{equality-2}) must hold for any operator $O$, one deduces that
\begin{equation}\label{TH1-bis}
    \rho=\mathbb{I} - \left( \mathsf{U} \rho \, \mathsf{U}^\dagger \right)^* \quad.
\end{equation}
So far in this proof, $\mathsf{U}$ is an arbitrary unitary matrix. In particular, in the case Eq.(\ref{U-def}) considered in the Main text, one straightforwardly obtains Eq.(\ref{TH1}).
Moreover, by recalling that the diagonal entries of the single particle density matrix are the expectation values of the particle density,  $\langle {n}_{j \alpha}\rangle=\rho_{j\alpha, j\alpha}$, and by taking the diagonal entries of Eq.(\ref{TH1}), one deduces that $2 \rho_{j\alpha, j\alpha}-1=0$\, $\forall j,\alpha$, which straightforwardly implies Eq.(\ref{TH2}). Finally, the statement (\ref{pinning}) of the theorem was proven in Ref.[\onlinecite{NJP-rossi-dolcini}].

In a similar manner one can prove Eq.(\ref{TH-chiral}) related to the implication of chiral symmetry on the single particle density matrix. The main difference with respect to the above proof is that, because of the anti-linear character of $\mathcal{S}$, in the first line of Eq.(\ref{equality-2}), one has to replace $O_{IJ} \rightarrow O^*_{IJ}=O_{JI}$. After substituting $\mathsf{U}$ with the matrix $\mathsf{U}_0$  entering the definition (\ref{S-def}) of chiral symmetry,   Eq.(\ref{TH-chiral}) is eventually obtained.

\section{Calculation of the density deviation in the  strongly dimerized limit}
\label{AppB}
Here we provide details about the calculation of the dynamical evolution of the site density resulting from the quench protocol (\ref{quench}) applied of the SSH model threaded by a flux $\varphi$. In particular, we shall demonstrate the asymptotic behavior Eq.(\ref{deltanA-asympt}). Exploiting the cell translational invariance one can write 
\begin{eqnarray}
{n}_\alpha(t) &\equiv & \frac{1}{L}\sum_{j}  \langle {n}_{j\alpha}\rangle = \frac{1}{L}\sum_{k}\langle {n}_{k\alpha}\rangle = \nonumber \\
&=&\frac{1}{L}\sum_{k} (\rho_{-})_{\alpha \alpha}(k,t) = \frac{1}{2}\pm \delta n(t) 
\end{eqnarray}
where the $\pm$ sign refers to  $\alpha=A,B$, respectively.
Recalling from Sec.\ref{sec4B} of the Main Text that the evolution of the $k$-bloch is $\rho_{-}(k,t)=[\sigma_0 -\hat{\mathbf{d}}(k,t)\cdot \boldsymbol\sigma]/2$ and that $\hat{\mathbf{d}}(k,t)$ is given by Eq.(\ref{hatd(k,t)}), one has  
\begin{eqnarray}
\delta n (t) &=& \frac{1}{2L} \sum_k (\hat{\mathbf{d}}^i(k) \times \hat{\mathbf{d}}^f(k))_z \, \sin [2 |\mathbf{d}^f(k)|\, \tau] = \nonumber \\
&=& \frac{1-r^2}{2L} \sum_{k =-\pi}^\pi 
        f(k+\frac{\varphi}{L},\tau;r)  \label{deltan-pre}  
        \end{eqnarray}
where $\tau$  is the dimensionless time given in Eq. (\ref{tau-def}), $r$ is the dimerization parameter, and
\begin{equation}\label{f(k)-def}
f(k,\tau;r)= \sin k   \frac{ \,\sin\big[2 \tau \sqrt{1+r^2+2 r \cos k } \big]}
        {1+r^2+2 r \cos k }    \quad.
\end{equation}
Here we have exploited Eq.(\ref{d-vec}) and the form (\ref{quench}) of the quench protocol, where the pre- and post-quench Hamiltonian exhibit the same single-particle spectrum [see Eq.(\ref{spectrum}) for $U=0$].
Because $f(k,\tau;r)$ is a $2\pi$-periodic function of its argument $k$, by exploiting  its Fourier series expansion  $f(k,\tau;r)=(2\pi)^{-1}\sum_{n=-\infty}^{+\infty} c_n(\tau;r) \exp[i k n]$,   some straightforward algebra enables one to rewrite
\begin{eqnarray}
\delta n(t) =    \frac{(1-r^2)}{4\pi} \sum_{m=-\infty}^{+\infty} c_{mL}(\tau;r)   e^{i m \varphi  } \quad, \label{deltan-pre2} 
\end{eqnarray}
where 
\begin{eqnarray}
\lefteqn{ c_{mL}(\tau;r) =  \int_{-\pi}^\pi dk \, f(k,\tau;r) \, e^{-i k mL } =} && \label{c-coeff} \\ &=&  \int_{-\pi}^\pi dk \,  \sin  k \frac{\sin\big[2\tau \sqrt{1+r^2+2r\cos k}  \big] }{1+r^2+2r\cos k }    \, e^{-i k m L}\nonumber
\end{eqnarray}
By noting  that $c_{-mL}=-c_{mL}$ and that
 $c_0=0$, one can rewrite Eq.(\ref{deltan-pre2}) as
 \begin{eqnarray}
\delta n(t) =  i  \frac{(1-r^2)}{2\pi} \sum_{m=1}^{+\infty} c_{mL}(\tau;r) \sin(   m\varphi )  \quad. \label{deltan-pre3} 
\end{eqnarray}
So far, no approximation has been made. We shall now take the limit of strong dimerization $r \ll 1$ and approximate the coefficients (\ref{c-coeff}) as
\begin{eqnarray}
   \lefteqn{ c_{mL}(\tau;r)  \simeq     \int_{-\pi}^\pi dk \,  \sin k     \,  \sin\big[2\tau (1+ r \cos k )\big] \, e^{-i\kappa mL}}   &&\nonumber\\ &  =&    \,  \frac{\pi mL}{2 \tau r} \,    i^{mL+1} J_{mL}(2\tau r)\left[  e^{i 2 \tau} +(-1)^{mL } e^{-i 2 \tau} \right] \, .\label{c-coeff-approx}
\end{eqnarray}
Assuming e.g. $L$ even and inserting Eq.(\ref{c-coeff-approx}) into Eq.(\ref{deltan-pre3}) one obtains\begin{eqnarray}
     \langle \delta \hat{n} \rangle  \simeq    
   -\sum_{m=1}^{+\infty}  \, (-1)^{\frac{ m L}{2}}   \,  \frac{m L}{2 \tau r} \,  J_{mL}( 2 \tau r)    \cos(2  \tau)   \sin(\varphi m ) \hspace{0.3cm}
\end{eqnarray}
In particular, in the time domain $2\tau r < 2L$,   the first term with $m=1$ is dominant over the other ones, and one finally obtains Eq.(\ref{deltanA-asympt}) that well captures the behavior of Fig.\ref{Fig2}.\\
 
\section{Maximally localized Wannier functions in a finite size ring}
\label{AppC}
In this Appendix we provide some details about the calculation of the MLWFs in a finite size ring-shaped lattice with cell translational invariance.
The construction given below is based on the assumption to deal with a many-particle state that is the  half-filled ground state of a two-band Hamiltonian, with a  completely filled   lower band and a  completely empty   upper band.

It is worth  emphasizing that this approach   is not limited to the initial ground state of the pre-quench Hamiltonian. Indeed, when a quench is performed in a system with cell translational invariance, the evolved many body state can always be regarded, at any given time $t$, as the half filled ground state of a fictitious  two-band  Hamiltonian $H(t)$, where time appears as a parameter.  
This is because, as pointed out in Sec.\ref{sec4B}, the dynamical evolution of the single particle density matrix is decoupled in $k$-space, $\rho(t)=\oplus_{k \in BZ}  \rho_{-}(k,t)$, where each 
evolved $k$-block can be
written as   the ground state $\rho_{-}(k,t)=   [ \sigma_0 - \hat{\mathbf{d}}(k,t)\cdot \boldsymbol{\sigma}  ]/2$  of a two-level system with Hamiltonian  $h(k,t)=\hat{\mathbf{d}}(k,t)\cdot \boldsymbol{\sigma}$, where $\hat{\mathbf{d}}(k,t)$ is the unit vector given in Eq.(\ref{hatd(k,t)}). Thus, the evolved 
 many-particle state at time $t$ is   the half filled ground state of the fictitious Hamiltonian $H(t)=\gamma \sum_k c^\dagger(k) [\hat{\mathbf{d}}(k,t)\cdot \boldsymbol\sigma] c(k)$, with two flat bands $E_\pm(k)=\pm \gamma$.  \\

Keeping in mind that the outline approach  holds time by time, in the remaining part of this Appendix we shall make our notation lighter and drop the time $t$. 
We define the   lower band Wannier state centered at the cell $m$,  in the Bloch gauge ${g}$, as 
\begin{equation} \label{Wannier-state}
    | m \,  -  \rangle[{g}] = \frac{1}{\sqrt{L}} \sum_k e^{- i k m} \, e^{i {g} (k)}  | k \,  - \rangle
\end{equation}
where, without any loss of generality, we have labelled the cells as $m=0, \, \ldots \, L-1$. Moreover, $| k \,  - \rangle = | k \rangle \otimes | u_-(k) \rangle$ represents the single particle lower band Bloch state   in an arbitrary  reference gauge, which we choose for definiteness as 
\begin{equation}
    | u_\pm(k) \rangle =   \left[ 2(1 \mp \hat{d}_z(k))\right]^{-1/2}  
    \begin{pmatrix}
        \hat{d}_x(k) - i \hat{d}_y(k) \\
        \pm 1 - \hat{d}_z(k)
    \end{pmatrix} \quad,
\end{equation}
where $\hat{d}_x(k), \, \hat{d}_y(k), \, \hat{d}_z(k)$ are the components of the  unit vector $\hat{\mathbf{d}}(k)$ given in Eq.(\ref{hatd(k,t)}) and describing the quench dynamics of the two-band model  (at time $t$). \\

The  Wannier wavefunction is thus obtained by projecting the Wannier state (\ref{Wannier-state}) on the real space state $\langle j \alpha|$, with $j=0,\ldots L-1$ denoting any cell and $\alpha=A,B$  the sublattice index, and by exploiting $\langle j \, \alpha  | k \,  -  \rangle=\exp[i k j] u_-^\alpha(k)/\sqrt{L}$. One obtains
\begin{eqnarray}
\psi_{m, -}^{{g}} (j, \alpha)&=&  \langle j \, \alpha  | m \,  - \rangle[ {g}]  = \nonumber \\
&=&\frac{1}{L} \sum_k e^{i k (j-m)} \, e^{i {g} (k)}  u_-^\alpha(k) \label{psi-Wannier}
\end{eqnarray}
and the corresponding probability of finding the particle in the $j$-th cell is given by:
\begin{equation}
    P_{m -}^{{g}}(j)=| \psi_{m -}^{{g}} (j, A) |^2 + | \psi_{m -}^{{g}} (j, B) |^2 \quad .
\end{equation}
Note that, by construction, the various Wannier wavefunctions (\ref{psi-Wannier}) labelled by $m$ are related to each other by a translation, namely $\psi_{m, -}^{{g}} (j, \alpha)=\psi_{0, -}^{{g}} (j-m, \alpha)$, so that it is sufficient to analyze the properties of the Wannier function $\psi_{0, -}$ labelled by the cell $m=0$. For this reason, we shall henceforth drop  the ``$0,-$" subscript, and simply redenote $\psi_{0, -}(j,\alpha)\rightarrow \psi(j,\alpha)$ and $P_{0, -}(j)\rightarrow P(j)$.

In particular, we now want to evaluate the expectation value and the variance of the position operator. In an chain lattice with OBCs or in an infinitely long system  the position operator is naturally defined as   the Hermitean operator $\hat{X}$ whose  one-body representation in the real space  is identified by the diagonal matrix $X=\text{diag}(1,\ldots,L-1,L)\otimes \sigma_0$, where $L$ is the number of cells. However, because   of the  ring-shaped geometry considered here, such operator would manifestly break the PBCs. It is thus useful to adopt a regularized position operator $\hat{E}=\exp[i  2\pi \hat{X}/{L}]$, which is not Hermitean, but whose one-body representation   in the site basis is  the diagonal matrix $E=\text{diag}(e^{i\frac{2\pi}{L}} , e^{i\frac{4\pi}{L}} \ldots,  1 ) \otimes \sigma_0$ that fulfills the PBCs.

If one computes the expectation value of $\hat{E}$ on the single particle Wannier function   $\psi(j,\alpha)$, one obtains a complex number
\begin{eqnarray}
    \langle e^{i \frac{2 \pi}{L} X} \rangle [{g}] &\equiv& \sum_{j} e^{i \frac{2 \pi}{L} j} P^{{g}}(j)  
     \equiv  \Omega[{g}] \, e^{i \Theta [{g}]} \label{R-Theta-def}
\end{eqnarray}
with modulus $\Omega[{g}]$ and phase $\Theta[{g}]$. Equation (\ref{R-Theta-def}) can now be interpreted in terms of directional statistics\cite{fisher-book}, i.e. as the result of a stochastic process where  points are randomly distributed on a unit circle according to the probability distribution $P^{{g}}(j)$, where $j=0, \, \ldots \, L-1$ identifies an ``angular" position $2\pi j/L$ along the circle, with the cell $L-1$ being the ``left" nearest neighbour of cell $0$. 
It is possible to show that
the phase   
$\Theta[{g}]$  can be interpreted as the center of the Wannier function, which  in the present case of a finite size system is   gauge dependent, while the modulus   $\Omega[{g}]$   in Eq.(\ref{R-Theta-def}) is closely related to the (gauge dependent)   spread of the Wannier function through $\Delta X=\sqrt{1-\Omega^2}\, L/2\pi$.  This correspondence becomes apparent when taking the thermodynamic limit,   as we shall see below. However, even for a finite size system,
one can
notice that, if the    distribution $P^{{g}}(j)$ is sharply peaked on one ``angle" $2\pi j/L$, then the phase $\Theta$ in Eq.(\ref{R-Theta-def})  represents such an angle and the modulus is maximal, $\Omega \simeq 1$. In contrast,
 if  $P^{{g}}(j)$ is uniform along the ring, then the modulus $\Omega$ is minimal, $\Omega \simeq 0$.

In order to determine the MLWFs, we have   to determine the gauge ${g}^*$ such that $\Omega[{g}^*]$ is maximal, or equivalently $\Omega^2[{g}^*]$ is maximal.  
It is possible to find an exact solution to the problem. Indeed we can rewrite Eq.(\ref{R-Theta-def}) as 
\begin{eqnarray}
 \lefteqn{  \Omega[{g}] \, e^{i \Theta [{g}]}  =   } & & \nonumber \\
    &=& \frac{1}{L^2} \sum_{j,\alpha}  \sum_{k,q}   e^{i(q+\delta k-k)j} e^{i[{g}(q)-{g}(k)]} (u_{-}^\alpha(k))^* u_{-}^\alpha(q) = \nonumber \\
    &=&\frac{1}{L} \sum_k \langle u_-^{g}(k+\delta k) | u_-^{g}(k) \rangle \label{R-Theta-def-2} 
\end{eqnarray}
where we have denoted   $| u_-^{g}(k) \rangle=e^{i {g}(k)} | u_-(k) \rangle$,  $\delta k=2\pi/L$, and we have exploited the Kronecker-$\delta$   obtained from the summation over $j$.
The quantity
$ \xi^{g}_{-}(k)=\langle u_-^{g}(k+\delta k) | u_-^{g}(k) \rangle$ is a complex number whose modulus is gauge independent (and smaller than $1$), while its phase critically depends on the gauge. In order to maximize $\Omega[{g}]$, one needs to choose the gauge ${g}^*$, in which  the phases of all $\xi^{g}(k)$ in Eq.(\ref{R-Theta-def-2}) are as equal as possible, so that   the terms  sum  as coherently as possible  over $k$.
Such optimal gauge choice can be determined by observing that  
\begin{equation}
    \varphi_{B-}= \sum_k \text{arg}\left\{ \xi_-^{g}(k) \right\} \quad \mod{2\pi} \, ,
\end{equation}
which can be regarded as the   discrete version of the Berry phase, is a gauge independent quantity. This becomes apparent by writing (${\rm mod} 2\pi$)
\begin{eqnarray}
   \lefteqn{ \varphi_{B-} = \sum_k \text{Im} \left( \ln \xi_-^{g}(k) \right)  =   \text{Im} \left(  \ln \prod_k \xi_-^{g}(k) \right)  = }   & & \nonumber \\
    &=& \text{Im} \left\{  \ln \Big[ \ldots \langle u_-^{g}(k+2\delta k) | u_-^{g}(k+\delta k) \rangle \times \right.   \nonumber \\
    & &  \left. \hspace{2cm} \times \langle u_-^{g}(k+\delta k) | u_-^{g}(k) \rangle \ldots \Big] \right\} \nonumber \\
    &=& \text{Im} \left\{  \ln \Big[ \text{tr} \prod_k \rho_-(k)  \Big] \right\} 
\end{eqnarray}
where $\rho_-(k)=|u_-^{g}(k)\rangle \langle u_-^{g}(k)|$ is a gauge invariant projector. Thus, one can  define the   gauge ${g}^*(k)$ as
\begin{equation}
    \text{arg}\left\{ \xi_-^{{g}^*}(k) \right\}=\text{Im} \left\{ \ln \xi_-^{{g}^*}(k) \right\}=\frac{\varphi_{B-}}{L} \quad \forall k
\end{equation}
which represents the gauge, in which the (discrete version of the) Berry connection  is constant. It is also possible to show that $\varphi_{B -}$ tends to the actual Berry phase when taking the thermodynamic limit.
 
 It is then straightforward to notice that  such a gauge  is characterized by the following  properties. The phase $\Theta$, which represents the angular center of the Wannier wavefunction, coincides with the discrete version of the Berry phase
 \begin{equation}
  \Theta[{g}^*]=\frac{1}{L} \varphi_{B-} \quad. 
 \end{equation} 
 whereas the
 modulus $\Omega$ takes the maximal possible value
 \begin{equation}
 \Omega[{g}^*]=  \underset{[ g]}{\rm max} \,\Omega[{g}]= \frac{1}{L} \sum_k |\xi^{{g}^*} (k)|= \frac{1}{L} \sum_k |\xi^{{g}} (k)| \quad.
 \end{equation}  
Furthermore, it is possible to show that, in the thermodynamic limit $L\rightarrow \infty$, i.e. $\delta k\rightarrow 0$, one has 
\begin{eqnarray}
    \Theta[{g}] \rightarrow \Theta   \equiv \left(\frac{2\pi}{L} \right) \frac{1}{2\pi} \int  dk \, \langle  u_-^{g}(k) |i \partial_k| u_-^{g}(k) \rangle \quad
\end{eqnarray}
i.e. the Wannier center becomes gauge independent and, when converted into spatial coordinates  
$\langle X \rangle = L \Theta/2\pi$ (in units of the lattice spacing), it coincides with the known result reported in Refs.[\onlinecite{vanderbild_PRB_1997,vanderbild_PRB_2002,vanderbild_RMP_2012}].
Moreover, the modulus $\Omega$   tends to 
\begin{eqnarray}
\Omega^2[{g}]& \rightarrow &  \Omega^2 \equiv 1-\left(\frac{2\pi}{L} \right)^2 \frac{1}{2\pi} \int  dk \,   \big| | \partial_k u_-^{g}(k) \rangle \big|^2 + \nonumber \\
    &&   +\left(\frac{2\pi}{L} \right)^2 \left[\frac{1}{2\pi} \int  dk \,  \langle  u_-^{g}(k) | i \partial_k | u_-^{g}(k) \rangle\right]^2  \quad,
    \end{eqnarray}
which implies that the Wannier spread  in spatial  coordinates reads  
\begin{eqnarray}
      \Delta X=\sqrt{\langle X^2 \rangle  -   \langle X \rangle^2}  =  \frac{L}{2\pi} \sqrt{1- \Omega^2 } \quad. \label{DeltaX-Omega}
    \end{eqnarray}
given in Ref.[\onlinecite{kohn_PR_1959,vanderbild_PRB_1997,vanderbild_PRB_2002,vanderbild_RMP_2012}].      It is thus evident that maximizing $\Omega$ corresponds to minimizing the spreading of the Wannier functions.

 In conclusion, it is possible to exactly find the  MLWFs even in a finite size 1D ring-shaped lattice. Such functions have a center of mass with linear coordinate $ \varphi_{B-}/2\pi$, where we have converted the angular position into the linear one through the factor $L/2\pi$.  Moreover  the angular spread  of the MLWFs can be identified from Eq.(\ref{DeltaX-Omega}) as 
\begin{eqnarray}
 \lambda =    \frac{2\pi}{L}  \,  \Delta X     =        \sqrt{ 1- \Omega^2[{g^*}]  } \quad.
\end{eqnarray}
This quantity is what we have plotted in Fig.\ref{Fig4}(c)
\section{Minimization of symmetry breaking quantifiers $Q_{glob}^C$ and $Q_{loc}^C$}
\label{AppD}
Here we discuss some technical details about the  
 phase set $\{ \theta\}$ that minimizes the charge symmetry breaking quantifiers  (\ref{QC-global-def}) and (\ref{QC-local-def}), in the various cases described in Fig.\ref{Fig5}. 
It turns out that there are two   types of sets minimizing the quantifiers $Q_{glob}^C$ and $Q_{loc}^C$. 
 The first one is  
\begin{equation}\label{theta-u-def}
\{\theta_u\} \equiv \begin{cases}
\theta_{j,A} =+\delta/2  \\
\theta_{j,B} =-\delta/2  \\
\end{cases} \quad \hspace{1cm} j=1,\ldots L  
\end{equation}
with $\delta$ denoting a staggering, while the second one is given by Eq.(\ref{theta-ell}). 

When the flux is absent ($\varphi=0$) and the $\mathcal{C}$-symmetry is broken locally by the on-site potential $U$,
the staggering value minimizing the quantifiers  is $\delta=0$, and the two $\{ \theta \}$ sets (\ref{theta-u-def}) and (\ref{theta-ell}) both coincide with $\{ \theta \} \equiv 0$. 
This holds for both  $Q^C_{glob}$ and $Q^C_{loc}$,  regardless of whether the ground state is metallic or insulating. It simply reflects the fact that a change in the gauge [see Eq.(\ref{gauge-def})] has no effect whatsoever on the on-site potential, whereas it can introduce a $\mathcal{C}$-symmetry breaking in the hopping terms.

When $U=0$ and the $\mathcal{C}$-symmetry is broken non locally by a flux $\varphi$, however, the two phase sets (\ref{theta-u-def}) and (\ref{theta-ell})  qualitatively differ, and one becomes more favorable with respect to the other.  In particular,  recalling Eqs.(\ref{gauge-def}) and (\ref{0-to-g}),   it is straightforward to see that the phase set (\ref{theta-u-def}) identifies a   transformation  to a gauge  where the flux    is described by a vector potential   uniformly distributed along all the links of the ring, as pictorially sketched  in Fig.\ref{Fig3}(a), with at most a staggering $\delta$ in alternating links. Such a set Eq.(\ref{theta-u-def}) turns out to be the most favorable   for the global quantifier $Q^{C}_{glob}$, for both the metallic and insulating states. Yet, the value of the small optimal    staggering     $\delta=O(1/L)$   depends on the dimerization parameter $r$  (with $\delta=0$ for the non-dimerized metallic case),  implying that the $\mathcal{C}$-symmetry is least broken when the vector potential has larger magnitude in the weaker links (say $w$) than in the stronger links (say $v$). 
In contrast, the set (\ref{theta-ell})  leads to the gauge pictorially  illustrated in Fig.\ref{Fig3}(b), where the vector potential is accumulated along the link located at $P_2$, i.e. on the opposite side of the ring with respect to the considered reference point $P^*$. This gauge turns out to be the most favorable for the local quantifier $Q^C_{loc}$, for both the metallic and the insulating states.
\begin{figure}[h]
\centering
\includegraphics[width= 8cm]{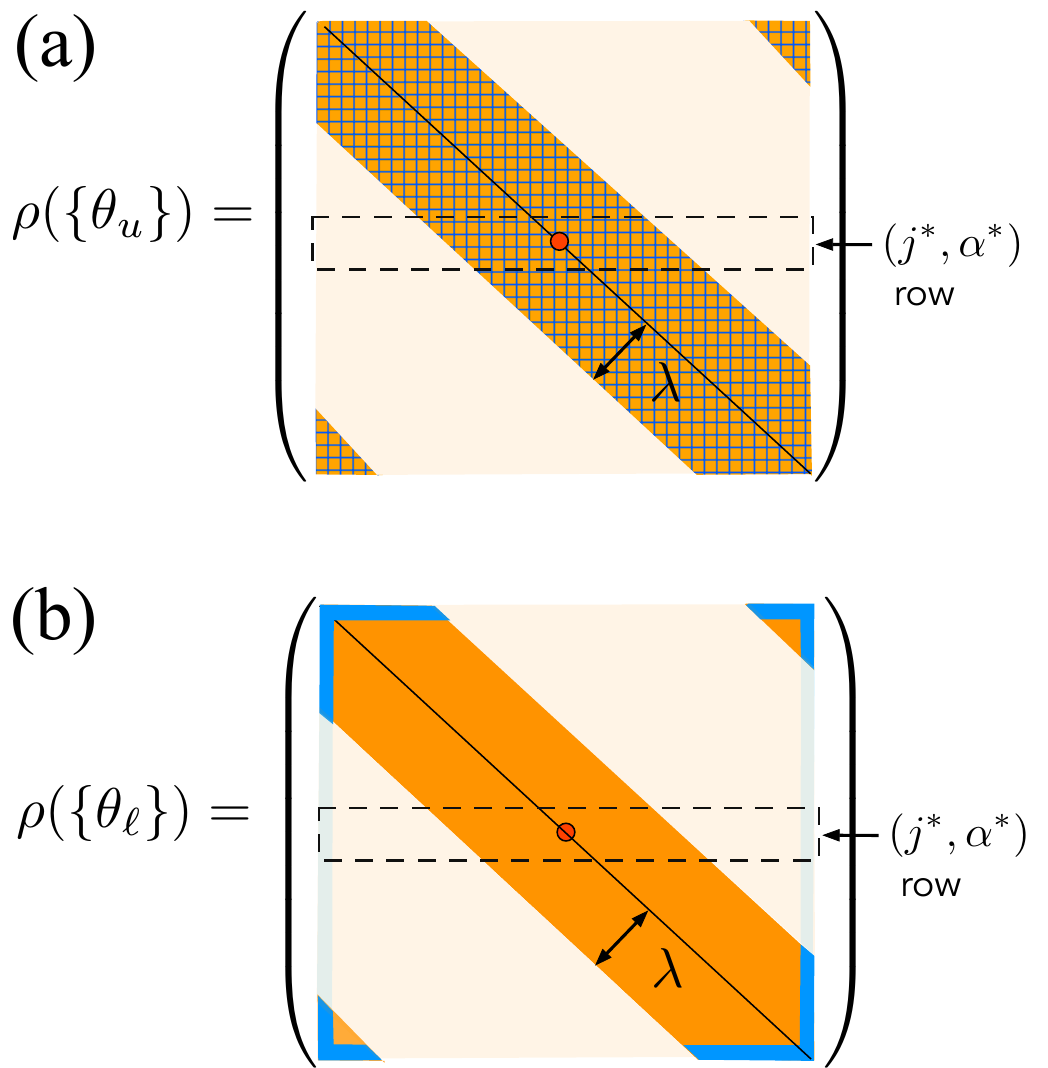}

\caption{\label{Fig8}  Sketch of the structure of the  single particle density matrix $\rho$  of the ground state of the SSH model with flux, evaluated in real space for two different $\{ \theta \}$ phase sets. The ivory areas identify the exponentially small entries of $\rho$,   while the cyan pattern highlights how  the flux phase $\varphi$ is distributed over the finite-value entries, here marked as the orange areas. (a) The case of the phase set $\{ \theta_u\}$ in Eq.(\ref{theta-u-def}),  corresponding to the gauge illustrated  in Fig.\ref{Fig3}(a): The flux is uniformly distributed over the various $\rho$ entries.   (b) The case of the phase set $\{ \theta_\ell\}$ in Eq.(\ref{theta-ell}), corresponding to the gauge illustrated  in Fig.\ref{Fig3}(b), where the flux phase is accumulated in the farthest link with respect to the considered reference point $P^*$. The dashed box highlights the $(j^*,\alpha^*)$ row, which determines $Q_{loc}^C$ in Eq.(\ref{QC-local-def}) and describes the correlations between the chosen reference point [specifically,  the site $P^*=(L/2, A)$, depicted as a red spot] and any other lattice site. An insulator   is locally insensitive to a flux   because in the in the local $\mathcal{C}$-breaking quantifier $Q_{loc}^C$,  the flux phase only appears in the  exponentially small tails located at the box   ends, since the localization length $\lambda$ is much smaller than the system size $L$.  }
\end{figure}

The different  impact of   the two phase sets $\{\theta_u\}$ and $\{\theta_\ell\}$   can now be illustrated by comparing the structure of  the single particle density matrix   $\rho(\{\theta\})$, when evaluated in the two corresponding gauges [see Eq.(\ref{rho-gauge})]. This is sketched in Fig.\ref{Fig8}, where the ivory areas correspond to exponentially small entries  of $\rho$,  the  blue areas describe how the flux phase $\varphi$ is distributed in its finite-valued entries, which are highlighted as orange areas and are arranged along a diagonal stripe, whose width is determined by the correlation length $\lambda$ of the quantum state. 
For the uniform set $\{\theta_u \}$ [see Fig.\ref{Fig8}(a)], the flux-phase $\varphi$  is uniformly distributed in all entries of $\rho$,  consistently with the   corresponding     gauge   in Fig.\ref{Fig3}(a).  In contrast, for the phase set $\{\theta_\ell\}$ [see Fig.\ref{Fig8}(b)],  the flux phase appears only along the frame of $\rho$, which corresponds precisely to the point $P_2$ located at the opposite side of the ring with respect to the reference point $P^*$, as also seen in the corresponding gauge   in Fig.\ref{Fig3}(b).

The global quantifier $Q_{glob}^C$ in Eq.(\ref{QC-global-def}) involves all entries of $\rho$, through the matrix $b^C$ in Eq.(\ref{bC-def}). In contrast, the   local quantifier $Q^C_{loc}$ in Eq.(\ref{QC-local-def}) picks up only the $(j^*,\alpha^*)$-th row of~$\rho$, here highlighted as a dashed box, which   contains the correlations between the reference point $P^*$   (red spot) and any other site $P$.
From Fig.\ref{Fig8} one can understand the different scaling laws obtained in Fig.\ref{Fig5}(b). Indeed, when the flux is absent, i.e. for a local $\mathcal{C}$-symmetry breaking induced by the staggered on-site potential, the   structure of $\rho$ is the same in both panels of Fig.\ref{Fig8}, and the quantifier depends only on correlation length $\lambda$, which can be identified with the localization length of the Wannier functions. While an insulator  exhibits a finite localization length, the Wannier functions in a metal are delocalized and the correlations extend   with a slow algebraic decay from the diagonal over the entire density matrix $\rho$, causing the weaker decay   $Q^C_{loc}\sim 1/\sqrt{L}$ obtained in the metallic case [red curve of Fig.\ref{Fig5}(b)], as compared to    $Q^C_{loc}\sim 1/{L}$ found in the insulating SSH case 
 [cyan curve of Fig.\ref{Fig5}(b)].

When the $\mathcal{C}$-symmetry is broken non locally by the flux~$\varphi$, however, the   structure of $\rho$  in Fig.\ref{Fig8}(a) and Fig.\ref{Fig8}(b) differ in the way the flux phase is distributed, and the $\{ \theta_\ell \}$  is the most favorable set for $Q_{loc}$. In the metallic state the correlation is quasi long ranged, i.e. the orange area extends with a power law decay over the entire row, up to the edges of the dashed box: The reference point $P^*$ can experience the presence of the $\mathcal{C}$-breaking flux, even if accumulated at the opposite ring side. This results   in the algebraic decay,    $Q^C_{loc}\sim 1/\sqrt{L}$ shown in the black curve of Fig.\ref{Fig5}(b). In contrast, 
because an insulator has a finite localization length $\lambda$, only sites located within such distance are correlated with $P^*$. The $\mathcal{C}$-breaking flux phase  is present   only at the  ends of the dashed box   in Fig.\ref{Fig8}(b), i.e. in entries that are exponentially small in the system size, leading to the exponential suppression $Q^C_{loc} \sim \exp[-L/\Lambda]$   found in the blue curve of Fig.\ref{Fig5}(b).  

Notice that, although the flux is effectively vanishing  at local level,   the global quantifier Eq.(\ref{QC-global-def}) involves {\it all} entries of the  single-particle density matrix $\rho$, including the blue corners in Fig.\ref{Fig8}(b) where the flux phase is accumulated. For this reason, the phase set $\{\theta_\ell \}$ is the most favorable for $Q^C_{loc}$, but unfavorable for $Q^C_{glob}$ as compared to $\{\theta_u\}$.   


\begin{thebibliography}{99}
\bibitem{note-on-cc} In order  to avoid  non-local effects that are trivially induced by non-local transformations, we shall   limit ourselves to the physically local  transformation (\ref{CC-def}) mapping particles into holes on the   same site, up to a possible phase factor.
\bibitem{nota-factor-2} Note that the phases in  Eqs.(\ref{CC-def}) and (\ref{gauge-def}) differ by exhibit a factor 2, due to the nature of $\mathcal{C}$-transformation, which  exchanges ${c}^\dagger \leftrightarrow {c}$.
\bibitem{nota-translational-symmetry} Note that the transformation  (\ref{gauge-def}) with the phase set Eq.(\ref{theta-ell})  would  break translational invariance.
\bibitem{nota-gaussian-state}
This is certainly true if the many-particle state is Gaussian,  $\boldsymbol\rho \propto \exp(-\sum_{I,J}{c}^\dagger_{I} M_{I,J} {c}^{}_J)$ with $M^\dagger=M$, a condition that is fulfilled in all the non-interacting models considered here. In strongly interacting systems, however, signatures of the $\mathcal{C}$-breaking may appear   in many-body observables only, while being absent in one-body observables.
\bibitem{nota-minimization} Taking into account  the cell translational invariance, the presence of dimerization and the flux, we have considered phase sets of the form $\theta_{j,A/B}=a \, j +\theta\pm \delta/2$, where  $\theta$, $a$ and $\delta$ are treated as variational parameters.
\end{thebibliography}
\end{document}